\documentstyle[12pt]{article}

\advance\textheight by \topskip
\begin{document}
%\baselineskip 1.1truecm
\tolerance=100000
%\input $disk1:[moretti.tek]feynman
%\input $disk1:[moretti.tek]macro
%
%%%%%%%%%%%%%%%%%%%%%%%%%%%%%% definitions
%%%%%%%%%%%%%%%%%%%%%%%%%%%%%% %%%%%%%%%%%%%%%%%%%%%%%%%%%%%%%%%%%%%
\def\Dir{\kern -6.4pt\Big{/}}%su lettere italiane minuscole
\def\DDir{\kern -7.6pt\Big{/}}%su lettere italiane maiuscole
\def\DGir{\kern -6.0pt\Big{/}}%su lettere greche
\def\Ord{\buildrel{\scriptscriptstyle <}\over{\scriptscriptstyle\sim}}
\def\simlt{\rlap{\lower 3.5 pt \hbox{$\mathchar \sim$}} \raise 1pt \hbox {$<$}}
\def\OOrd{\buildrel{\scriptscriptstyle >}\over{\scriptscriptstyle\sim}}
\def\simgt{\rlap{\lower 3.5 pt \hbox{$\mathchar \sim$}} \raise 1pt \hbox {$>$}}
\def\pl #1 #2 #3 {{\it Phys.~Lett.} {\bf#1} (#2) #3}
\def\np #1 #2 #3 {{\it Nucl.~Phys.} {\bf#1} (#2) #3}
\def\zp #1 #2 #3 {{\it Z.~Phys.} {\bf#1} (#2) #3}
\def\pr #1 #2 #3 {{\it Phys.~Rev.} {\bf#1} (#2) #3}
\def\prep #1 #2 #3 {{\it Phys.~Rep.} {\bf#1} (#2) #3}
\def\prl #1 #2 #3 {{\it Phys.~Rev.~Lett.} {\bf#1} (#2) #3}
\def\mpl #1 #2 #3 {{\it Mod.~Phys.~Lett.} {\bf#1} (#2) #3}
\def\rmp #1 #2 #3 {{\it Rev. Mod. Phys.} {\bf#1} (#2) #3}
\def\xx #1 #2 #3 {{\bf#1}, (#2) #3}
\def\preprint{{\it preprint}}
\def\sm{\ifmmode{{\cal {SM}}}\else{${\cal {SM}}$}\fi}
\def\mssm{\ifmmode{{\cal {MSSM}}}\else{${\cal {MSSM}}$}\fi}
\def\MH{\ifmmode{{M_{H^0}}}\else{${M_{H^0}}$}\fi}
\def\Mh{\ifmmode{{M_{h^0}}}\else{${M_{h^0}}$}\fi}
\def\MA{\ifmmode{{M_{A^0}}}\else{${M_{A^0}}$}\fi}
\def\MHpm{\ifmmode{{M_{H^\pm}}}\else{${M_{H^\pm}}$}\fi}
\def\tb{\ifmmode{\tan\beta}\else{$\tan\beta$}\fi}
\def\ctb{\ifmmode{\cot\beta}\else{$\cot\beta$}\fi}
\def\ta{\ifmmode{\tan\alpha}\else{$\tan\alpha$}\fi}
\def\cta{\ifmmode{\cot\alpha}\else{$\cot\alpha$}\fi}
\def\tba{\ifmmode{\tan\beta=1.5}\else{$\tan\beta=1.5$}\fi}
\def\tbb{\ifmmode{\tan\beta=30}\else{$\tan\beta=30$}\fi}
\def\cab{\ifmmode{c_{\alpha\beta}}\else{$c_{\alpha\beta}$}\fi}
\def\sab{\ifmmode{s_{\alpha\beta}}\else{$s_{\alpha\beta}$}\fi}
\def\cba{\ifmmode{c_{\beta\alpha}}\else{$c_{\beta\alpha}$}\fi}
\def\sba{\ifmmode{s_{\beta\alpha}}\else{$s_{\beta\alpha}$}\fi}
\def\ca{\ifmmode{c_{\alpha}}\else{$c_{\alpha}$}\fi}
\def\sa{\ifmmode{s_{\alpha}}\else{$s_{\alpha}$}\fi}
\def\cb{\ifmmode{c_{\beta}}\else{$c_{\beta}$}\fi}
\def\sb{\ifmmode{s_{\beta}}\else{$s_{\beta}$}\fi}
\def\be{\begin{equation}}
\def\ene{\end{equation}}
\def\ba{\begin{eqnarray}}
\def\ena{\end{eqnarray}}
\def\ar{\rightarrow}
\def\F{\ifmmode{\cal F}\else{$\cal F$}\fi}
\def\X{\ifmmode{\cal X}\else{$\cal X$}\fi}
\def\Y{\ifmmode{\cal Y}\else{$\cal Y$}\fi}
\def\Z{\ifmmode{\cal Z}\else{$\cal Z$}\fi}
\def\li{\ifmmode{p_i,\lambda}\else{$p_i,\lambda$}\fi}
\def\lj{\ifmmode{p_j,\lambda'}\else{$p_j,\lambda'$}\fi}
\def\l #1{\ifmmode{p_{#1},\lambda_{#1}}\else{$p_{#1},\lambda_{#1}$}\fi}
\def\m #1{\ifmmode{q_{#1},\lambda_{#1}}\else{$q_{#1},\lambda_{#1}$}\fi}
\def\r #1{\ifmmode{r_{#1},-}\else{$r_{#1},-$}\fi}
%%%%%%%%%%%%%%%%%%%%%%%%%%%%%%%%%%%%%%%%%%%%%%%%%%%%%%%%%%%%%%%%%%%%%%
%%%%%%%%%%
\thispagestyle{empty}
\setcounter{page}{0}

\begin{flushright}
{\large DFTT 26/94}\\
{\large DTP/94/46}\\
{\rm December 1994\hspace*{.5 truecm}}\\
{\rm Revised July 1995\hspace*{.5 truecm}}\\
\end{flushright}

\vspace*{\fill}

\begin{center}
{\Large \bf Intermediate mass standard model Higgs boson at the proposed
CERN LEP$\otimes$LHC
$ep$ collider\footnote{Work supported in part by Ministero
dell' Universit\`a e della Ricerca Scientifica (SM) and by a
University of Durham
Studentship and a World Lab. Fellowship ICSC (GAL).}}\\[2.cm]
{\large Ghadir Abu Leil$^a$ and Stefano Moretti$^{a,b,}$\footnote{Address
after September 1995: Cavendish Laboratory,
University of Cambridge,
Madingley Road,
Cambridge, CB3 0HE, U.K.} }\\[2cm]
{\it a) Department of Physics, University of Durham,}\\
{\it    South Road, Durham DH1 3LE, U.K.}\\[1cm]
{\it b) Dipartimento di Fisica Teorica, Universit\`a di Torino,}\\
{\it    and INFN, Sezione di Torino,}\\
{\it    V. Pietro Giuria 1, 10125 Torino, Italy.}\\[2cm]
\end{center}
\vspace*{\fill}

\begin{abstract}
{\normalsize
\noindent
The production of the \sm\ Higgs $\phi$ with intermediate
mass at the proposed CERN  LEP$\otimes$LHC $ep$ collider in
$\gamma q(\bar q)\rightarrow W^\pm\phi q'(\bar q')$,
$\gamma q(\bar q)\rightarrow Z^0\phi q(\bar q)$ and
$g\gamma\rightarrow q\bar q\phi$ events is studied. This is
done for all possible (massive) flavours of the quarks $q(q')$ and
 using photons generated via Compton back--scattering of laser
light. We study signatures in which
the Higgs decays to $b\bar b$--pairs and the electroweak vector bosons
$W^\pm$ and $Z^0$ decay either hadronically or leptonically.
All possible backgrounds to these signals are also computed.
Flavour identification on $b$--jets is assumed.
Explicit formulae for the helicity amplitudes
of the above processes are given.}
\end{abstract}

\vspace*{\fill}

\newpage
%\baselineskip 1.1truecm
\subsection*{Introduction}

The Higgs sector is one of
the most investigated parts of the Standard Model (\sm)
\cite{sm,higgs},
yet is continues to be very elusive.
 So far the Higgs particle  has evaded all
searches. Nevertheless, a  lower limit on the mass of the \sm\ Higgs
$\phi$ of $\approx60~{\rm {GeV}}$ was extracted from the lack of $e^+e^-\ar
Z^0\ar Z^{0*} \phi$ events at LEP I \cite{limSM}. An upper bound of
$\approx1$ TeV is expected. This was derived by requiring the validity of
perturbation theory \cite{perturbativity} and the unitarity of the model
\cite{unitarity}.
Therefore, if the \sm\ Higgs $\phi$ exists,
we could expect it to be discovered by the next generation
of CERN high energy colliders: LEP II ($\sqrt s_{ee}=$160--200 GeV)
\cite{LepII} and the LHC ($\sqrt s_{pp}=10, 14$ TeV)\cite{LHC}.\par
LEP II will be able to cover the mass range $M_{\phi} < $80--100 GeV.
A Higgs with a larger mass should be searched for
at the LHC. At LEP II $\phi$ can be detected\footnote{And produced
via the Bjorken bremsstrahlung process  $e^+e^-\ar
Z^{0*}\ar Z^{0} \phi$ \cite{bremSM}.}
through a large variety of decay channels, the most favoured being
$Z^{0} \phi\ar(\mu^{+}\mu^{-})(b\bar b)$.
A Higgs boson
with mass $M_\phi\OOrd 130$ GeV is clearly detectable at the LHC using
the four--lepton mode
$\phi\rightarrow Z^0Z^0\rightarrow \ell\bar\ell\ell\bar\ell$\footnote{With
$\phi$ produced via $gg$-- \cite{gg} or $W^\pm W^\mp$-- and
$Z^0Z^0$--fusion \cite{WWZZ}.}. Due to the QCD backgrounds typical of
hadron colliders,
it is still controversial whether it is possible
to detect an intermediate mass Higgs\footnote{Via the associated production
with a $W^\pm$ boson (decaying leptonically to $\ell\nu$) \cite{gny,wh}
or a $t\bar t$ pair (with one $t$ decaying semileptonically to
$b\ell\nu$) \cite{rwnz,tth}.}
in the mass range 90 $\Ord M_{\phi}
\Ord 130$ GeV (where $\phi$ mainly
decays to $b\bar b$ pairs). In this mass range $\phi$ can be
searched for through the rare
$\gamma\gamma$ decay mode and this relies on the fact that both a high
luminosity and
a very high di--photon mass resolution must be achieved at
the LHC \cite{gamgam}.
It is also unclear whether it is possible to cleanly detect the intermediate
\sm\ Higgs in
the $\phi\rightarrow b\bar b$ channel using the $b$--tagging capabilities of
vertex detectors \cite{btagg,SDC}. The main difficulties being the
 expected low signal rates after reconstruction, the necessity to have
an accurate control on all the possible background sources and to
achieve a very high
$b$--tagging performance \cite{last}.\par
In the distant future,
cleaner environments for studying the Higgs boson parameters
will be the $e^+e^-$ linear accelerators ($\sqrt s_{ee}=350-2000$ GeV)
\cite{guide,NLC,ee500,LC92,JLC}.\par
At the Next Linear Collider (NLC), with $\sqrt s_{ee}=300-500$ GeV
\cite{LC92},
the Higgs boson can be searched for through a large number of channels
over the whole intermediate mass range \cite{BCDKZ}. The dominant
production mechanism is the Bjorken reaction for
$\sqrt s_{ee}$ below 500 GeV while
the $W^\pm W^\pm$-- and $Z^0Z^0$--fusion processes
\cite{fusionSM} will dominate at larger energies.
At $\sqrt s_{ee}\OOrd500$ GeV \cite{ee500}
a heavy Higgs can be detected in the four--jet modes
$\phi\rightarrow W^\pm W^\mp,Z^0Z^0\rightarrow jjjj$ \cite{BCKP,4jet}
in addition to the $4\ell$--mode.
At higher energies, $\sqrt s_{ee}=1-2$ TeV \cite{JLC},
the same search strategies still hold with the fusion mechanisms becoming the
dominant ones.\par
  The conversion of the linear
$e^+e^-$ NLCs into $\gamma\gamma$ and/or $e\gamma$ colliders, by
photons generated via Compton back--scattering of laser light,
will provide new possibilities for detecting and studying the Higgs
boson \cite{laser}. In $\gamma\gamma$ collisions two of the important
channels will be: the production of a heavy Higgs
(up to $\approx 350$ GeV)
by a triangular loop of heavy fermions or $W^\pm$,
with the detection via the decay mode $\phi\rightarrow
Z^{0}Z^{0}\rightarrow q\bar q\ell^+\ell^-$ at  $\sqrt s_{ee}=500$ GeV
\cite{heavyphphSM}, and the
process $\gamma\gamma\rightarrow
t\bar t \phi$, which appears more useful than the corresponding $e^+e^-$ one
in measuring the top Yukawa coupling $t\phi$, at $\sqrt s_{ee}=1-2$ TeV
\cite{Yukawa}.
The $e\gamma$ option at linear colliders can be exploited for studying
Higgs production via the process $e\gamma\rightarrow \nu_eW\phi$,
at $\sqrt s_{ee}=1-2$ TeV and
over the mass range 60 GeV $\Ord M_\phi\Ord$ 150 GeV \cite{Boos,Hagiwara},
using the signature $W^-\phi\rightarrow (jj)(b\bar b)$  \cite{Cheung}.
The cross section for the above process at such $\sqrt s_{ee}$'s
is comparable to the fusion process
$e^+e^-\rightarrow \bar\nu_e\nu_eW^{\pm*} W^{\mp*}\rightarrow
\bar\nu_e\nu_e\phi$
and larger than the bremsstrahlung reaction
$e^+e^-\rightarrow Z^{0*}\rightarrow Z^0\phi$.
Finally, it has been shown in ref.~\cite{Eboli} that the process
$e\gamma\rightarrow e\gamma\gamma\rightarrow e\phi$ is the most important
mechanism for $\phi$--production
at $\sqrt s_{ee}=500$ GeV, for $M_\phi\OOrd 140$ GeV.\par
Let us now consider the production of the \sm\ Higgs boson at $ep$ machines.
This seems to be beyond the capabilities of HERA \cite{epHERA}, which
has been primarily designed for providing accurate data on the
proton structure functions in the small--$x$ region, more than for
Higgs searches \cite{GGv}.
In the future, another $ep$ collider is contemplated,
the CERN LEP$\otimes$LHC accelerator: it will combine an
electron/positron beam from LEP II and a proton beam from the
LHC \cite{LHC,epLEPLHC}. A detailed study on the detectability
of an intermediate mass \sm\ Higgs boson at such a machine has been presented
in ref.~\cite{ZeppenfeldLHC}. This is based on the $W^\pm W^\mp$-- and
$Z^0Z^0$--fusion processes \cite{GGv,WWZZep,corrWWep}, with $\phi$ decaying
to $b\bar b$. It has been shown that it should be possible to detect
$\phi$ provided that a high luminosity and/or an
excellent $b$--flavour identification can be achieved. Only recently
has the possibility of resorting to back--scattered laser photons at
the $ep$ CERN collider been suggested \cite{CheungLEPLHC},
searching for, e.g., $\gamma q\ar
q'W^\pm\phi$ events, with $\phi\ar b\bar b$ and $W^\pm\ar \ell\nu$ or $jj$,
which should give detectable Higgs signals if  good $M_{b\bar b}$ invariant
mass resolution can be achieved and efficient $b$--tagging can be
performed.\par
  The purpose of this paper is to study the following reactions
at the LEP$\otimes$LHC $ep$ collider
\be\label{proc1}
q\gamma\ar q'W^\pm\phi,
\ene
\be\label{proc2}
q\gamma\ar q Z^0\phi,
\ene
\be\label{proc3}
g\gamma\ar q\bar q\phi,
\ene
in the intermediate mass range
of $\phi$, for all possible (anti)flavours of the (anti)quarks $q(q')$,
using laser back--scattered photons.
We discuss their relevance to the detection of the \sm\ Higgs
and the study of its parameters, with the Higgs decaying to $b \bar b$--pairs
and assuming flavour identification on its decay products.\par
Although process (\ref{proc1}) has already been studied
in \cite{CheungLEPLHC}, and the part of the analysis devoted to it here
largely overlaps that study, we decided nevertheless to include it
for completeness and since, in principle,
we can slightly improve the results previously obtained. In fact,
since we consider
heavy quarks we include additional Higgs bremsstrahlung off quarks
in the amplitudes, even though these are suppressed with respect
to contributions coming from diagrams involving $\phi W^+W^-$ vertices.
 We also computed all the necessary rates for all
the relevant backgrounds exactly, whereas these latter contributions were only
estimated in \cite{CheungLEPLHC}.
Reaction (\ref{proc3}) has been analysed in \cite{qqhHERA} for
\mssm\ neutral Higgses, $b$--quarks and using bremsstrahlung photons
but to our knowledge, neither the larger energy option
available at LEP$\otimes$LHC nor the possibility of using laser
back--scattered photons has been exploited.\par
 There are at least two important motivations for analysing processes
(\ref{proc1})--(\ref{proc3}) at the LEP$\otimes$LHC collider.
First, if the \sm\ Higgs boson turns out to have an intermediate mass
greater than the maximum value that can be reached by LEP II
and if the LHC detectors are not able to achieve the necessary
performances for the predicted Higgs measurements \cite{last}, the $ep$ CERN
collider
will be the first alternative option
available for studying such a Higgs, as it will certainly be
operating before any NLC.
Second, although both the cross sections and the luminosity at
LEP$\otimes$LHC are expected to be small if compared with the LHC ones,
the CERN  $ep$ option will
constitute the first TeV energy environment partially free from the
enormous QCD background typical of purely hadronic colliders.
Moreover, processes (\ref{proc1})--(\ref{proc3}) have the advantage,
compared
to the $W^\pm W^\mp$-- and $Z^0Z^0$--fusion mechanisms, that
the additional heavy particles $W^\pm$ and $Z^0$ (and also $t$, in principle)
can be used for tagging purposes by searching for their decays,
thus increasing the signal to background ratio.\par
The plan of the paper is as follows.
In Section II we give details of the calculation
and the numerical values adopted for the various parameters.
Section III is devoted to the discussion of the results, while the
conclusions are in Section IV. The helicity amplitudes for
 processes (\ref{proc1})--(\ref{proc3}) are presented in the Appendix.\par

\subsection*{Calculation}

Fig.~1 shows all the Feynman diagrams at tree level contributing to
the reactions (\ref{proc1}) and (\ref{proc2}) in
the unitary gauge, where $(q,q',V)$ represent the possible
combinations $(d,u,W^-)$, $(u,d,W^+)$ and
$(q,q,Z^0)$ respectively (in the case of  process (\ref{proc2}) only the
first eight diagrams of fig.~1 contribute). Fig.~2 shows the Feynman
diagrams at tree level for process (\ref{proc3}).
All quarks have been considered massive, so
diagrams with a direct coupling of $\phi$
to the fermion lines have been taken into account.\par
The amplitudes squared have been computed by means of the
 spinor techniques of refs.~\cite{ks,mana} and, as a check, also
by the method of ref. \cite{hz}
%\footnote{We do not present here the
%corresponding helicity amplitudes, since they can be obtained by
%appropriate changes of couplings as a subset of those to be given in
%\cite{galMSSM} for the case of \mssm\ Higgs bosons.}.
The matrix elements for the processes
$\bar d\gamma\ar \bar u W^+\phi \  / \  \bar u\gamma \ar \bar dW^-\phi$ and
$\bar q\gamma\ar \bar qZ^0\phi$
can easily be obtained by trivial operations
of charge--conjugation.
All of the above amplitudes have been tested for gauge invariance.
We were also able to ``roughly''\footnote{See
footnote 9 below.} reproduce, with appropriate couplings,
hadron distributions
and luminosity function of the photons, the results
of ref.~\cite{CheungLEPLHC} and of ref.~\cite{qqhHERA}.
Moreover, since a simple adaptation of the implemented formulae
(by changing photon couplings from quarks into leptons and setting the
quark masses equal to zero) allowed us to reproduce the computation
of ref.~\cite{Cheung}, we have checked our helicity amplitudes in this
way also.\par
As proton structure functions we adopted the HMRS set B \cite{HMRS}
(this was done in order to make comparisons with already published
work easier),
setting the energy scale equal to the center--of--mass (CM) energy
  at the parton level
(i.e. $\mu=\sqrt{\mathaccent 94{s}}_{\rm {parton}}$).
The strong coupling constant $\alpha_s$, which appears in the
gluon initiated
processes, has been evaluated at two loops, for $\Lambda_{QCD}=190$
MeV, with a number $N_f=5$ of active
flavours and a scale $\mu$ equal to that used for the proton
structure functions.
We are confident that
changing the energy scale and/or distribution function choice should not
affect our results by more than a factor of two\footnote{We verified
this in few cases by comparing the actual results
to the ones obtained from the more recent set of structure functions
MRS(A) \cite{MRSA}.}.\par
For the energy spectrum of the back--scattered (unpolarized) photon
we have used \cite{backscattered}
\begin{equation}\label{backsc}
F_{\gamma/e} (x)= \frac{1}{D(\xi)}\left[1-x+\frac{1}{1-x}-\frac{4x}{\xi(1-x)}
      +\frac{4x^2}{\xi^2(1-x)^2}\right],
\end{equation}
where $D(\xi)$ is the normalisation factor
\begin{equation}
D(\xi)=\left(1-\frac{4}{\xi}-\frac{8}{\xi^2}\right)\ln(1+\xi)
+\frac{1}{2}+\frac{8}{\xi}-\frac{1}{2(1+\xi)^2},
\end{equation}
and $\xi=4E_0\omega_0/m_e^2$, $\omega_0$ is the incoming
laser photon energy
and $E_0$ the (unpolarized) electron/positron energy. In eq.~(\ref{backsc})
$x=\omega/E_0$ is the fraction
of the energy of the incident electron/positron carried by the
back--scattered photon,
with a maximum value
\begin{equation}
x_{\rm {max}}=\frac{\xi}{1+\xi}.
\end{equation}
In order to maximise $\omega$ while avoiding $e^+e^-$ pair creation, one takes
$\omega_0$ such that $\xi=2(1+\sqrt 2)$ and one gets the typical
values $\xi\simeq 4.8$, $x_{\rm {max}}\simeq 0.83$, $D(\xi)\simeq 1.8$.\par
In the case of $q(g)\gamma$ scattering from $ep$ collisions,
the total cross section $\sigma$ is obtained by folding
the subprocess cross section $\hat\sigma$ with
the photon $F_{\gamma/e}$ and hadron $F_{q(g)/p}$ luminosities:
\begin{equation}
\sigma(s_{ep})=
\int_{x^\gamma_{\rm {min}}}^{x^\gamma_{\rm {max}}}dx^\gamma
\int_{x^{q(g)}_{\rm {min}}}^{1-x^\gamma}dx^{q(g)}
F_{\gamma/e}(x^\gamma)
F_{q(g)/p}(x^{q(g)})
\hat\sigma(\hat s_{q(g)\gamma}=x^\gamma x^{q(g)}s_{ep}),
\end{equation}
where $\hat s_{q(g)\gamma}$ is the CM
energy at parton (i.e., $q(g)\gamma$) level, while
\begin{equation}
x^\gamma_{\rm {min}}x^{q(g)}_{\rm {min}}=
\frac {(M_{\rm {final}})^2}{s_{ep}},
\end{equation}
where $M_{\rm {final}}$ is the sum of the final state particle masses.\par
The multidimensional integrations have been performed numerically
using the Monte Carlo
routine VEGAS \cite{vegas}.\par
To our knowledge, a detailed study, as for the cases of $e\gamma$ and
$\gamma\gamma$ collisions \cite{backscattered}, on the efficiency of the laser
back--scattering method in converting
$e\ar \gamma$ at $ep$ colliders does not exist. In this paper we
assume for the
effective $\gamma p$ luminosity the same as the $ep$ one, therefore
the conversion efficiency of electrons into backscattered $\gamma$'s
is one.
For the discussion of the results we have adopted an overall
total integrated luminosity ${\cal L}=3$ fb$^{-1}$ per year, the value
of ref.~\cite{CheungLEPLHC}.\par
  For the numerical part of our work, we have taken
$\alpha_{em}= 1/128$ and  $\sin^2\theta_W\equiv s^2_W=0.23$, while
for the gauge boson masses and widths:
$M_{Z^0}=91.175$ GeV, $\Gamma_{Z^0}=2.5$ GeV,
$M_{W^\pm}=M_{Z^0}\cos\theta_W\equiv M_{Z^0}c_W$ and
$\Gamma_{W^\pm}=2.2$ GeV.
For the fermions we have: $m_e=0.511\times10^{-3}$ GeV,
$m_\mu=0.105$ GeV, $m_\tau=1.78$ GeV, $m_u=8.0\times10^{-3}$ GeV,
$m_d=15.0\times10^{-3}$ GeV, $m_s=0.3$ GeV, $m_c=1.7$ GeV, $m_b=5.0$ GeV and
$m_t=175$ GeV \cite{CDF}, with all widths equal to zero apart from
$\Gamma_t\approx1.58$ GeV, adopting its tree--level expression).
All neutrinos have been considered
massless: i.e.,
$m_{\nu_e}=m_{\nu_\mu}=m_{\nu_\tau}=0$.
The Branching Ratios (BRs) of the Higgs boson were extracted from
ref.~\cite{BRs}.\par
We have analysed the processes (\ref{proc1})--(\ref{proc3}) over the mass
range 60 GeV $\Ord M_{\phi}\Ord$ 140 GeV
and for $ep$ CM energy ranging from 0.5 to 3.0 TeV, with  special
attention devoted to the case $\sqrt s_{ep}=1.36$ TeV, corresponding
to the collision of an electron/positron beam from LEP II
and a proton beam from LHC \cite{CheungLEPLHC}.\par

\subsection*{Results}

In figs.~3--5 we present the dependence of processes
(\ref{proc1})--(\ref{proc3}) on the collider CM energy,
for a selection of Higgs masses: $M_{\phi}=60, 80, 100, 120$ and 140 GeV.
Summations over all possible combinations of (anti)flavours have been
performed (the top contributions in the final states are
included\footnote{As a first
approximation only combinations of two flavours within the same
quark doublet have been computed for
process (\ref{proc1}), setting all Cabibbo--Kobayashi--Maskawa terms
equal to one.}), as
well as the integration
over the initial $g/q(\bar q)$-- and $\gamma$--structure
functions.
A general feature in  figs.~3 and 5 is the
rapid increase of all the plots with $\sqrt s_{ep}$, especially for
$\sqrt s_{ep}\OOrd 1$ TeV.
This is because for $\sqrt s_{ep}$ much larger than the final particle masses,
phase space effects are quantitatively unimportant.
 The same effect is less
evident in fig.~4, since process (\ref{proc2}) is affected by the $s$--channel
structure of the corresponding
Feynman diagrams, whereas [part of ] these are in $t$--channel
for process [(\ref{proc1})](\ref{proc3}).
 We also notice that the cross section for the
process $ep\ar W^\pm\phi X$ is much larger than  that of $ep\ar Z^0\phi X$.
This is due to
two reasons:
first, the coupling $\phi W^+W^-$ is larger than $\phi Z^0Z^0$ and second,
in process (\ref{proc1}) there are additional diagrams (i.e., \# 9--12 in
fig.~1),
some of which (i.e., \# 11 and 12) are not suppressed by Yukawa couplings.\par
In Tab.~I we give the cross sections at the
LEP$\otimes$LHC CM energy $\sqrt s_{ep}=1.36$ TeV.
To show the importance of
the relative contributions of the various flavours entering in the
subprocesses (\ref{proc1})--(\ref{proc3}), we give their separate rates
in Tab.~II at  $M_{\phi}=60$ GeV.
For reaction
(\ref{proc1}) at a fixed $\sqrt s_{ep}$,
increasing the Higgs mass reduces the top quark contributions, this
is due to  the limited phase space available, while the
 light flavours contributions (i.e., $q=u,d,s,c$ and $b$) do not
change significantly.
For example,
the top contribution to process (\ref{proc1}) diminishes
from 1.4\% to 0.12\% when $M_\phi$ increases from 60
to 140 GeV, whereas the contributions from $up$ ($down$)
[$strange$] \{$charm$\}--initiated
processes vary from $\approx53$ (35) [8] \{3\}\%
to $\approx$ 64 (29) [5] \{2\}\%. For process (\ref{proc2}) there is no
substantial phase space effect of this kind, since we cannot have
top contributions here.
Thus the numbers do not differ as much: they are
$\approx 74$ (16) [4] \{5\} $<0.6>$ \%  to  $\approx80$
 (14) [3] \{3\} $<0.33>$\%,
 with the numbers in the ``brackets'' $<>$ corresponding to $b$--contributions.
For reaction (\ref{proc3}), things change dramatically because, on the one
hand, top--lines are not connected to the initial state
as in (\ref{proc1}) and the phase space suppression
due to the large top mass is important only if
$\sqrt s_{ep}\Ord 1$ TeV, and on the other hand,
the Higgs always couples to the very massive top--quark through the
($\sim m_t$) Yukawa coupling, in all Feynman diagrams at tree--level.
Because of this $\sim m_q$ coupling the very light flavours $q=u,d$
and $s$ give here completely
negligible contributions, while $c$-- and $b$--fractions are suppressed
by a factor of $\approx(m_t/m_c)^2\approx10^4$ and
$\approx(m_t/m_b)^2\approx1225$,  with respect to the top ones.
Therefore, for process (\ref{proc3}), the top--contribution is by far
the dominant one for $\sqrt s_{ep}\OOrd 1$ TeV and all
$\phi$--masses\footnote{Whereas for
$\sqrt s_{ep}\Ord 1$ TeV the $c$--contribution is the largest one:
in this case the effect of the
$q\gamma $ electromagnetic coupling,
which favours
$c$--quarks, is dominant on the Yukawa $q\phi$ electroweak one, which favours
$b$--quarks.}. The corresponding numbers at the LEP$\otimes$LHC
energy, varying $M_\phi$ in the range $60-140$
GeV,
are: $\approx$0.0016--0.0013\% for $u$--, $\approx$0.0013--0.0011\% for $d$--,
$\approx$0.29--0.28\% for $s$--,
$\approx$17--20\% for $c$--, $\approx$14--21\% for $b$-- and
$\approx$69--58\% for $t$--quarks.\par
Next, we checked if neglecting
diagrams 1--6 [and 9--10] of process (\ref{proc2})[(\ref{proc1})] inside the
matrix elements, as done in ref.~\cite{CheungLEPLHC},
where all quark masses were set equal to zero, could be a source of
error\footnote{We expect differences
coming from phase space  effects to be negligible for
the light flavours $u,d,s,c$
and $b$, since $m_q<<\sqrt s_{ep}$ for all of them.}. In doing this we needed
to apply some cuts to avoid collinear and soft singularities
(in the couplings of the incoming photon to
the outgoing quark $q_{\rm out}$) that would otherwise make our amplitudes
divergent. To do this, we require, e.g.,
$|\cos\theta_{\gamma q_{\rm out}}|<0.95$
and $|p_{q_{\rm out}}|>3$ GeV: restrictions which are reasonably
compatible with eventual
requirements from the detectors\footnote{Since similar cuts
were not listed in ref.~\cite{CheungLEPLHC},
we were unable to reproduce exactly the numbers
there computed.}. Setting again $\sqrt s_{ep}=1.36$ TeV
and $M_\phi=60$ GeV, we have found percentage differences only of the order
of 1 in 1000 in the case of light flavour final states, and of
$\approx$2\% for the contribution $b\gamma\ar tW^-\phi$
+ c.c., in
process (\ref{proc1}). For reaction (\ref{proc2}), differences are appreciable
only in the case of $c$-- and $b$--quarks, these being $\approx$3\%
and
$\approx$13\%,
respectively. These mass effects are approximately the same over the whole
intermediate $M_{\phi}$ range. However, due to the relative flavour
contributions
of Tabs.~IIa--IIb, when one sums over all of these
the effects are largely
washed out. We also notice
that the errors due to neglecting the quark masses are larger for process
(\ref{proc2}) than for (\ref{proc1}),
since in the latter there are also contributions (dominant with
respect to the
Higgs bremsstrahlung) coming from $\gamma\ar W^+W^-$
splitting whereas at tree--level there is no corresponding
$\gamma\ar Z^0Z^0$ coupling.
Obviously, taking into account the masses in process (\ref{proc3}) is
crucial, since there the Higgs is always produced through the
Yukawa couplings $q\phi$.\par
We know that in the mass range 60 GeV $\Ord M_\phi\Ord$ 140 GeV the
dominant Higgs decay mode is $\phi\ar b\bar b$. The corresponding
BR in the above interval varies
from $\approx0.85$ at $M_\phi=60$ GeV to
$\approx0.38$ at $M_\phi=140$ GeV, where the off--shell $W^{\pm *}W^\mp$
decay channel begins to be competitive \cite{BRs}. So, in order to maximise
the number of signal events we look for the $\phi\ar b\bar b$ signature.
We further require
flavour identification of $b$--jets, exploiting the possibilities offered by
$b$--tagging techniques, to reduce the large QCD backgrounds.\par
In processes (\ref{proc1})--(\ref{proc3}) we have additional decaying
particles\footnote{In principle, we also have
$t$--quarks in process (\ref{proc1}) which could decay
to $bW$--pairs, but in practise, contributions involving top quarks
are here generally
quite small if compared to those of the other flavours and substantially
negligible when we sum up all different combinations.}:
a $W^\pm$ in $q\gamma\ar q'W^\pm\phi$, a $Z^0$ in
$q\gamma\ar q Z^0\phi$ and two $t$'s in the $g\gamma \ar t\bar t\phi$
contribution.
So we expect the following
possible final signatures\footnote{We know that
in all processes (\ref{proc1})--(\ref{proc3}) we can have additional
$b$'s from $t/Z^0$--decays or $b\gamma/g\gamma$--fusion,
but we assume that complications coming from the fact of taking in those events
a wrong combination $b\bar b$ can be largely avoided if we restrict to keep
$b\bar b$--invariant masses in the window $|M_{b\bar b}-M_\phi|<5$ GeV
(see later on).}:
$$
ep\ar W^\pm\phi X\ar (\ell\nu_\ell)(b\bar b)X,
$$
\be\label{lepton}
ep\ar Z^0\phi X\ar (\ell\bar\ell)(b\bar b)X,
\ene
or
$$
ep\ar W^\pm\phi X\ar (jj)(b\bar b)X,
$$
\be\label{jet}
ep\ar Z^0\phi X\ar (jj)(b\bar b)X,
\ene
(where $X$ represents the untagged particles in the final states)
depending on whether the electroweak
massive vector bosons decay leptonically or hadronically,
respectively\footnote{We do not exploit here possible missing energy
decays $Z^0\ar \nu\bar\nu$ in process (\ref{proc2}).}.
As for process (\ref{proc3}) we expect the signature
\be
ep\ar q\bar q\phi X\ar jj(b\bar b)X
\ene
for light quark contributions, and
\be\label{nosign}
ep\ar t\bar t\phi X\ar b\bar b W^\pm(b\bar b)X
\ene
for top--quarks (with BR$(t\ar bW)\approx1$).
\par
Therefore out of the $\approx56-22[6-0.6]$ initial femtobarns of reaction
(\ref{proc1})[(\ref{proc2})] at $\sqrt s_{ep}=1.36$ TeV
and for $M_\phi=60-140$ GeV,
assuming ${\cal L}=3$ fb$^{-1}$,
we expect  $\approx99-18$[$11-<1$] events for hadronic decays,
and $\approx42-8$[$2-<1$] for leptonic modes, whereas for
reaction (\ref{proc3}), starting from $\approx3.8-0.24$ fb, we end up with
$\approx10-<1$ events (7 of these come from $t\bar t\phi$ production with
$M_\phi=60$ GeV), per year.\par
The irreducible backgrounds to the above signatures are
$ep\ar W^\pm Z^0 X\ar W^\pm (b\bar b)X$ and
$ep\ar \bar tbX\ar b\bar bW^\pm X$ for process (\ref{proc1}),
$ep \ar Z^0 Z^0 X\ar Z^0 (b\bar b)X$
for (\ref{proc2}), and $ep\ar q\bar q Z^0X\ar q\bar q (b\bar b)X$
for (\ref{proc3}). These are always present, independently of
the $W^\pm/Z^0$ decay modes in processes (\ref{proc1})--(\ref{proc2}).
In addition, multi--jet photoproduction, $W^\pm$ + jets,
$Z^0$ + jets  and $t\bar tX\ar b\bar b W^\pm X$
production and decay  events must
be also considered.\par
A few remarks concerning the $t\bar bX$ background are needed here.
We have mentioned earlier that we take the $e\ar \gamma$
conversion efficiency $\epsilon$ (into back--scattered photons) equal
to 1, which implies that all the incoming
electrons are converted into photons and hence removed from the
interaction site.
This motivates us to consider $\gamma p$ initiated processes only, and
not $ep$ ones.
Single--top production proceeds in $\gamma p$ collisions
through the partonic subprocesses $q\gamma\ar
q'W^{\pm *}\gamma\ar q't\bar b+q'\bar tb\ar q' b\bar b W^\pm$ (i.e.,
via $\gamma W^\pm$--fusion)
and $g\gamma\ar tbW^\pm\ar b\bar b W^+ W^-$ (i.e., via $\gamma\ar
W^+W^-$ splitting and $g\gamma$--fusion), whereas in $ep$ collisions it
happens via $e^-g\ar \nu_eW^{\pm *}g\ar t\bar b\nu_e$. While this latter
process has a very large cross section (approximately 1200 fb at
$\sqrt s_{ep}=1.36$ TeV), the sum of the first
two gives rates generally at the level of one order of magnitude larger than
the ones of the signal $qW^\pm\phi\ar qW^\pm(b\bar b)$,
for $m_t=175$ GeV
(see below). Therefore,
we would like to stress that it is extremely important
that an efficiency $\epsilon$
greater than $\approx 90\%$ should be achievable, otherwise
a  non--negligible fraction $(1-\epsilon)$
of the single--top background proceeding
via $W^\pm g$--fusion would enter in the experimental sample, inducing
a strong suppression of the signal versus background ratio. In fact,
the production rate of the $q W^\pm\phi$ signal via bremsstrahlung
photons
is more than ten times smaller than the one via backscattered $\gamma$'s
\cite{CheungLEPLHC}.\par
While $b$--tagging
identification should drastically reduce the
backgrounds where $b$--quarks are not present in the final states,
this requirement is not generally enough if they are. In this case,
one has to look for invariant masses of the $b\bar b$--pair in a
window around $M_\phi$, since the most part of the signals
lie within this region. In the case of top--resonant backgrounds
(i.e., $\bar t bX$ and $t\bar tX$) we can also exploit the cut, e.g.,
$|M_{bW\ar bjj}-m_t|>15$ GeV, which should be very effective in reducing
hadronic $W^\pm$--decays
since top--peaks are quite narrow (in fact,
$\Gamma_t\approx1.58$ GeV for
$m_t=175$ GeV).
Finally, if the Higgs mass turns out to be close
to the $Z^0$--mass, the precise absolute normalizations of
the processes involving $M_{b\bar b}$ resonances are
needed.\par
Assuming good $b$--tagging performances
such that it is possible
to drastically eliminate the non--$b$
multi--jet photoproduction, $W^\pm$ + jets and
$Z^0$ + jets  background events \cite{CheungLEPLHC}, and
that the $M_{b\bar b}$ cut is sufficient to suppress
the above processes in the case of $\gamma^* / g^*\ar b\bar b$ splitting,
we end up having to deal only with the backgrounds
$ep\ar W^\pm Z^0 X\ar W^\pm (b\bar b)X$,
$ep\ar \bar tbX\ar b\bar bW^\pm X$,
$ep \ar Z^0 Z^0 X\ar Z^0 (b\bar b)X$,
$ep\ar t\bar tX\ar b\bar b W^\pm X$
and $ep\ar q\bar q Z^0X\ar q\bar q (b\bar b)X$.
Moreover, we should  not forget that an additional drastic rejection
factor on the multi--jet background
comes from requiring that $M_{jj}/M_{\ell\bar\nu_\ell,\ell\bar\ell}$
has to reproduce
$M_{W^\pm}$ or $M_{Z^0}$ for processes (\ref{proc1})--(\ref{proc2}), and
 that $M_{bW\ar bjj} \approx m_{t}$ for (\ref{proc3}) when $q=t$
(since this  flavour is by far the largest partonic contribution
at the LEP$\otimes$LHC energy).\par
In order to study the background rates,
we have implemented
their matrix elements in {\tt {FORTRAN}} codes generated by
MadGraph \cite{Tim} and HELAS \cite{helas}\footnote{Since
process $ep\ar t\bar tX \ar
b\bar b W^\pm X$ was already
studied in ref.~\cite{Baur}, we also checked that the helicities
amplitudes we
obtained reproduce the results of that paper
(for bremsstrahlung photons).}. The total
cross sections of these processes are displayed in Tab.~III, at $\sqrt s_{ep}
=1.36$ TeV, for the same $\gamma$-- and $g/q(\bar q)$--structure functions
and parameters employed for the signal processes. We notice that backgrounds
are in general much larger than the corresponding signals, both
for the top--resonant cases (continuum backgrounds) and for the
$Z^0\ar b\bar b$ ones (discrete backgrounds). While in the former case
this happens because of the top--resonant peaks, in the latter
we have that the $qZ^0$ coupling does not depend on the $q$--mass (contrary
to the Higgs one),
so light quarks give large contributions here. This is especially evident
in the case of the reaction $ep\ar q\bar qZ^0$. The rates for $ep\ar Z^0Z^0 X$
are of the same order of magnitude as the signal $ep\ar Z^0\phi X$:
in this case the contributions from $Z^0$--bremsstrahlung
off quarks in the background (we do not have triple vector boson vertices
in this case) are comparable to those of the signal in which $\phi$ is emitted
from a $Z^0$--line.\par
However, in principle these very large rates should not be
a problem since processes $ep\ar W^\pm Z^0 X$, $ep \ar Z^0 Z^0 X$ and
$ep\ar q\bar q Z^0X$ are really important only when $M_\phi\approx M_{Z^0}$,
whereas $ep\ar \bar tbX\ar b\bar b W^\pm X$ and
$ep\ar t\bar t X\ar b\bar b W^\pm X$ are highly reduced when applying
a cut in the $b\bar b$--invariant mass (i.e., $M_{b\bar b}\approx
M_\phi$) and eventually, for
$W^\pm$--hadronic decays, also the cut $M_{bW}\approx m_t$ can be used.
In fig.~6 we give the differential
distributions in the invariant mass $M_{b\bar b}$
for those backgrounds in which the $b\bar b$--pair does not come from a
$Z^0$--resonance: i.e., $\bar tb X\ar b\bar b W^\pm X$ and
$t\bar tX\ar b\bar b W^\pm X$ ($W^\pm$--BRs are not included).
For backgrounds containing a $Z^0\ar b\bar b$ resonance, we naively assume
that all the $M_{b\bar b}$ spectrum is contained in the region
$|M_{b\bar b}-M_{Z^0}|\leq2\Gamma_{Z^0}=5$ GeV.\par
Since we are concentrating on $b\bar b$--invariant masses in
the $M_\phi$--region, we require
that $M_{b\bar b}$ of all events is in the window
$|M_{b\bar b}-M_{\phi}|<5$ GeV, assuming that 10 GeV will be
 the mass resolution of the detectors.
The fractions of the total cross sections
from $\bar tbX$-- and $t\bar tX$--production which pass
this cut are given by the
area under the $M_{b\bar b}$ distributions of fig.~6 between
$M_\phi -5$ and $M_\phi +5$  GeV, while we assume that those of the
 $Z^0$--resonant $b\bar b$--events are given by the
formula \cite{Cheung}
\be\label{peak}
\delta\sigma(Z^0)=\sigma(Z^0)\frac{{\rm max}(0, 10~{\rm GeV}-
|M_\phi-M_{Z^0}|)}{10~{\rm GeV}}.
\ene
In using the above equation we tacitly assumed that the $\phi\ar
b\bar b$ peaks
are also all contained in a region of 10 GeV around the
$\phi$--pole\footnote{In fact, the Higgs width at $M_\phi=140$ GeV is
$\Gamma_\phi\approx 0.01$ GeV.}.
The number of signal ($S$) and background ($B$) events and
their statistical significance ($S/\sqrt B$) are given in Tab.~IV,
for the three processes (\ref{proc1})--(\ref{proc3}) and the sum of
their backgrounds
separately, for the usual selection of $\phi$--masses, after the $M_{b\bar
b}$ cut. BRs of
hadronic and leptonic $W^\pm/Z^0$--decays,
giving the signatures in
eqs.~(\ref{lepton})--(\ref{nosign}),  are included both for processes
(\ref{proc1})--(\ref{proc2}) and
for the backgrounds. We do not make
any assumption about the $W^\pm$--decays
when $q=t$ in process (\ref{proc3}) and on the second $W^\pm$
in $tbX$ and $t\bar t X$, treating them
completely inclusively (i.e., such that $W^\pm$'s can decay
either hadronically or leptonically).\par
If, as criteria for the observability of a signal, we require a rate
$S\geq6$ events with a significance $S/\sqrt B>4$
for the detection of an isolated Higgs peak, while
for the case of Higgs peaks overlapping with $Z^0$ peaks
we require $S\geq10$ with $S/\sqrt B>6$ \cite{Cheung}, then we see from
Tab.~IV that the situation seems to be discouraging
both for hadronic
and leptonic $W^\pm$-- and $Z^0$--decays, if $M_{\phi} \OOrd 80$ GeV. It
does not look much better if one tries to make an
``inclusive'' analysis, summing the rates for signals and
backgrounds, as done in Tab.~V. This happens
because the largest signal
(i.e., $W^\pm\phi X$) has a huge background, whereas the other two
signals (i.e., $Z^0\phi X$ and $q\bar q \phi X$), even though
virtually free from backgrounds,
give very few events.\par
Therefore,  in the case of overlapping peaks it does not appear to be
any possibility  to disentangle the signals (see Tabs.~I
and III),  even after a few years of
running. However, if $|M_{\phi}-M_{Z^0}|\OOrd5$ GeV, region where only the
continuum backgrounds are effective, one can
exploit (in the case of hadronic $W^\pm$--decays)
the restriction $|M_{bW\ar bjj}-m_{t}|>15$ GeV (for
both the combinations $bW^+$ and $\bar bW^+$, assuming
to tag the positive gauge boson). For this, in fig.~7 we
plot the
differential distributions in $M_{b\bar b}$  of the $\bar tbX$ and
$t\bar t X$ backgrounds, after applying  the above $M_{bW}$ cut.
It is clear then how
this cut turns out to be extremely useful in rejecting the continuum
backgrounds, since their rates are now reduced of
$\approx81\%$ (for $tbX$) and of $\approx97\%$ (for
$t\bar tX$).
If we insert these reduction factors in Tabs.~IV--V the scenario changes
completely, since we have now to divide all $B$'s by a factor
of $\approx 13$, and multiply
all $S/\sqrt B$'s by $\approx\sqrt{13}$.
This gives significancies much
larger than 4 over almost all the intermediate Higgs mass range
($M_\phi\Ord120$ GeV). At the same
time, the
reduction factor for $W^\pm \phi X$ is just a few percent, since the
corresponding distribution in $M_{bW}$ is nearly flat (see
fig.~8): e.g., approximately $7\%$ for $M_\phi=60$ GeV
and $8\%$ for $M_\phi=140$ GeV.\par
%%%%%%%%%%%%%%%%%%%%%%%%%%%
A few comments concerning the mass resolution,
$|M_{\phi}-M_{b\bar b}|<5$ GeV, that we have used throughout
this paper are worth mentioning at this point.
In ref.~\cite{ZeppenfeldLHC},  a larger value was adopted. Here,
the fact that we perform the analysis at the parton level would enable
us to use for consistency a value of
$\approx7$ GeV \cite{last}
(which corresponds to a resolution between $11$ and 12.3 GeV
at the jet--level).
In addition, it has to be remembered
that the real performances of a possible
$ep$ CERN collider are not predictable at the moment, and it is not
inconceivable that by the time the LEP$\otimes$LHC machine
comes into operation further progresses in resolving
the mass spectra can be achieved. Therefore,
for the time being, we deliberately chose a smaller and more optimistic
value.
However, if eventually it turns out that such a performance
will not be feasible, the rates for a worse mass resolution (say 10 GeV)
can be readily deduced from the ones given here.
In fact, for the signals, due to the small Higgs
width in the intermediate mass range, they remain practically
unchanged. For the case of the discrete backgrounds we expect
smaller significancies only in the region around the $Z^0$ peak,
where it is already impossible to disentangle Higgs signals for
a mass resolution of 5 GeV.
Finally, for the continuum backgrounds, the numbers would be roughly a
factor two bigger (see figs.~6--7). Therefore,
an additional (overall) reduction
factor of $\approx\sqrt 2$ is expected, which should be compensated
for by a year of extra running at the same luminosity,
with respect to the case of higher mass resolution.\par
%%%%%%%%%%%%%%%%%%%%%%%%%%%%%%
So far we have assumed a 100\% acceptance and detection
efficiencies for $j/\ell$'s in the final states, the same
for $b$--tagging. This is obviously completely unrealistic, and before drawing
definite conclusions a full analysis (including kinematical cuts,
detector efficiencies, hadronization effects, etc ...) should be done.
We adopt here the set of kinematical cuts given in ref.~\cite{ZeppenfeldLHC}.
As the substantial part of Higgs signals would come
from reaction (\ref{proc1}), which furthermore
is the most affected by competitive backgrounds (contrary to
processes  (\ref{proc2})--(\ref{proc3}), which are
virtually free from backgrounds in the region where
Higgs signals can be disentangled, $M_{\phi}\ne M_{Z^0}$), we perform
the study for the case $W^\pm\phi X$
and for the corresponding (continuum) backgrounds.
\par
If we assume as
acceptance region the one
defined by\footnote{In the case of
the signal no requirement is imposed on the spectator jet from the
$q'$--quark in
(\ref{proc1}).}:
\begin{itemize}
\item transverse momentum $p_T^i$ of at least 20 GeV;
\item pseudorapidity $|\eta_i |$ less than 4.5;
\item separation $\Delta
R_{ij}=\sqrt{\Delta\eta_{ij}+\Delta\varphi_{ij}}>1$;
\end{itemize}
for all the $i$--th and $j$--th $b$'s and jets of the signature
$b\bar b jj X$, then the reduction factors for the $W^\pm\phi X$
signal and the $tbX$ and $t\bar tX$
background rates are: $R\approx16-7$ (for $M_\phi=60-140$ GeV),
 $\approx14$ and $\approx11$, respectively.
\par
That means that, on the one hand, the number of events
is reduced to a few units per year (from $\approx3$ at $M_\phi=140$
GeV to $\approx8$ at
$M_\phi=60$ GeV, for hadronic $W^\pm$--decays) whereas, on the other hand,
the effect on the significancies is a
reduction factor approximately equal to $4(2)$, for  $M_\phi=60(140)$
GeV. Therefore, we would conclude that
even though these selection criteria act in the direction of favouring
the backgrounds, largely spoiling the effectiveness of the $M_{bW\ar
bjj}$ cut, nevertheless, the final values we obtain for $S$,
$B$ and $S/\sqrt{B}$
shouldn't substantially modify
our above positive conclusions, but only imposing the
requirement of accumulating a  higher luminosity (in at least
two years time), in order to clearly
disentangle Higgs signals. In general, we would
like to stress here that our choice of kinematical cuts
could well be different from the one which will be at the end imposed
by the real LEP$\otimes$LHC detectors.
At present, in fact, the acceptances of these latter have yet not been
looked into, as even the most recent and complete studies
on the argument only deal
with simulations done for the LHC (see the
ATLAS {\cite{ATLAS} and CMS \cite{CMS}
Technical Proposals). That is,
we wonder if detectors designed for a $pp$ machine
will be the same and/or will work in
the same configuration even when they will be set up around a different
kind of machine, an $ep$ collider. Nothing prevent us then from thinking
that by the time the LEP$\otimes$LHC collider will be operating
both the improvement in the detection techniques and the necessity to
design the detectors in view of their best performances at an $ep$
machine, could end up by reducing the impact of the acceptance cuts on
the event selection procedure.
\par
%In general,
%we expect the acceptance cuts not to have a decisive
%role.\par
Concerning flavour
identification, it is clear that high $b$--tagging performances and excellent
not--$b$ rejection are needed, at least as the ones
expected at the LHC in the $pp$ mode \cite{last}.\par
Before concluding, we notice here how processes like
(\ref{proc1})--(\ref{proc3}) could turn out to be extremely interesting if
one considers their counterparts, e.g., in the
Minimal Supersymmetric Standard Model (\mssm). Here quark--Higgs couplings
proportional to $\tan\beta$ can enhance the signals up to ${\cal O}(1000)$
times for very large $\tan\beta$. This drastic enhancement happens
when considering
the contribution of diagrams involving the bremsstrahlung of the
pseudoscalar boson $A^0$ off massive
$down$--type quarks (i.e., $b$--quarks: hence masses should be
included). This occurs in all the Feynman diagrams of process
(\ref{proc3}), while it only happens for the suppressed graphs 1--6
[and 9,10] in (\ref{proc2})[(\ref{proc1})]. These latter contribute to the
total rate at the level of \% for the \sm\ case but are the only
surviving ones for the \mssm\ (since the pseudoscalar boson $A^0$
does not couple to vector
bosons at tree--level).
In addition, in processes (\ref{proc1})--(\ref{proc3}), once we
substitute $\phi$ by one of the \mssm\ neutral Higgses
$H^0$, $h^0$ and  $A^0$ and  we  also include the flavour changing cases
in which $\phi\leftrightarrow H^\pm$ and double Higgs productions in $q\gamma$
fusion ($W^\pm\leftrightarrow H^\pm$ and $Z^0\leftrightarrow H^0, h^0, A^0$),
we will have a very rich laboratory where all the fundamental
interactions of the \mssm\ can be carefully studied.
A complete analysis within this Model will be presented elsewhere
\cite{galMSSM}.

\subsection*{Summary and conclusions}

In summary, we have studied
the production cross sections of the \sm\ Higgs $\phi$ with mass in the
range 60 GeV $\Ord M_\phi\Ord$ 140 GeV at a next--generation $ep$
collider, with 500 GeV $\Ord\sqrt s_{ep}\Ord$ 3 TeV, through
the partonic processes
$$\gamma q(\bar q)\rightarrow q'(\bar q') W^\pm\phi,$$
$$\gamma q(\bar q)\rightarrow q(\bar q) Z^0\phi $$
and
$$
g\gamma\rightarrow q\bar q\phi,
$$
for all possible (massive) flavours of the quarks $q(q')$,
with incoming photons generated via Compton back--scattering of
laser light.\par
Special attention has been devoted to the case of the planned
CERN LEP$\otimes$LHC $ep$ collider (with $\sqrt s_{ep}\approx1.36$ TeV),
where signatures in which the Higgs decays to $b\bar b$--pairs were
studied, exploiting the possibilities given by $b$--tagging
techniques.\par
We concluded that at this machine, apart from the case
$M_\phi\approx M_{Z^0}$ which  is impossible to disentangle,
Higgs signals should be detectable above all the possible backgrounds
over the most part of the remaining intermediate mass range, by searching for
the hadronic decays of $W^\pm$'s
in process (\ref{proc1}), in particular after approximately a couple
of years
running at the luminosity ${\cal L}=3$ fb$^{-1}$ if $M_{\phi} \Ord 120$ GeV.
Due to the fact that the leptonic
decay channels of the $W^\pm$'s give small rates and that a cut in the
invariant mass $M_{bW}$ is not applicable in this case, no
possibility of detections exists if
$W^\pm\ar\ell\bar\nu_\ell$.
Therefore, in this respect, we disagree with the conclusions
given in ref.~\cite{CheungLEPLHC}.
In the case of processes
(\ref{proc2})--(\ref{proc3}), after the acceptance cuts here adopted
we expect to
get significant number of events only for a value of ${\cal L}$ much
bigger than the one assumed here (more than an order of magnitude).\par
In general, if the LHC detectors will not be able to achieve the necessary
performances for all the foreseen Higgs measurements, then the
LEP$\otimes$LHC
collider option could provide interesting prospects of
studying the \sm\ Higgs boson parameters (i.e., $M_\phi$, $\Gamma_\phi$,
BRs, etc ...) in the intermediate mass range, in an environment partially
free from the QCD background typical of $pp/p\bar p$
accelerators, especially if
larger $b$--tagging performances and/or a high luminosity
can be achieved, in advance of a possible future NLC.
\subsection*{Acknowledgments}

We are grateful to J.B.~Tausk for interesting discussions
and useful suggestions, to T.~Stelzer for his helpful advice on
using MadGraph and to W.J.~Stirling for reading the manuscript.

\subsection*{Appendix}

In this section we present the explicit formulae for the helicity amplitudes
of the processes we have studied. Definitions of $S$, $Y$ and $Z$ functions
and of other quantities ($p$, $\lambda$, $\mu$,
$\eta$, etc...), which enter in the following,
can be found in ref.~\cite{ioPR}, with identical notation.\par
Here, we introduce the definitions:
\begin{equation}
-b_1=-b_2=b_3=2b_4=2b_5=2b_6=2b_7=1
\end{equation}
for the coefficients of the incoming/outgoing four--momenta,
\begin{equation}
D_{V}(p)={1\over {p^2-M_V^2}},
\quad\quad
D_{q}(p)={1\over {p^2-m_q^2}}
\end{equation}
for the propagators, where $V=W^\pm,Z^0$  and $q=u$ or $d$,
\begin{equation}
N_i=[4(p_i\cdot q_i)]^{-1/2},\quad\quad\quad i=1,2
\end{equation}
for the gluon ($i=1$) and photon ($i=2$) normalisation factor,
where $p_i$($q_i$) is the massless vector four--momentum(any
four--vector not proportional
to $p_i$), with $i=1,2$ \cite{ks}.
The symbols $r_1$ and $r_2$ represent two light--like four--momenta
satisfying the relations
\begin{equation}
r_1^2=r_2^2=0,\quad\quad\quad r_1^\mu+r_2^\mu=p^\mu_4,
\end{equation}
($d\Omega_{r_1(r_2)}$ indicates the solid angle of $r_{1(2)}$ in the
rest frame of $p_4$)
\cite{ks}, $p_6$ and $p_7$ are antispinor four--momenta such that
\begin{equation}
p_6^\mu\equiv p_4^\mu, \quad\quad\quad p_7^\mu\equiv p_5^\mu,
\end{equation}
and
\begin{equation}
\sum_{\lambda=\pm} u(\li)\bar u(\li)=
{p\Dir}_i - m_i,\quad\quad{\rm {with}}~i=6,7,
\end{equation}
while
\begin{equation}
\sum_{\lambda=\pm} u(\li)\bar u(\li)=
{p\Dir}_i + m_i,\quad\quad{\rm {with}}~i=4,5.
\end{equation}
We also define the spinor functions\footnote{Throughout
this appendix we adopt the symbol ${\{\lambda\}}$ to denote a set
of helicities of all external particles in a given reaction,
$\sum_{\{\lambda\}}$ to indicate the usual sum over all their
possible combinations, and the symbol $\sum_{i=j,k,l,...}$
to indicate a sum over $j,k,l,...$ with index $i$.}
$$
{\X}_2=\sum_{\lambda=\pm}\sum_{i=1,3}(-b_i)
Y([2];[i];1,1)
Y([i];(2);1,1),
$$
$$
{\X}_4=\sum_{\lambda=\pm}\sum_{i=5,7}b_i
Y(\{1\};[i];1,1)
Y([i];\{2\};1,1),
$$
$$
{\X}^{qV(')}_{31}=
\sum_{\lambda=\pm}\sum_{i=4,6(5,7)}b_i
Y([3];[i];1,1)
Y([i];[1];c^q_{R_V},c^q_{L_V}),
$$
$$
{\Y}^{(')}_2=\sum_{\lambda=\pm}
\sum_{i=4,6(5,7)}b_i
Y([2];[i];1,1)
Y([i];(2);1,1),
$$
$$
{\Y}_4=\sum_{\lambda=\pm}
Y(\{1\};p_2,\lambda;1,1)
Y(p_2,\lambda;\{2\};1,1),
$$
$$
{\cal F}^{qV}_{31}=
\mu_1\eta_1 Y([3];[1];c^q_{L_V},c^q_{R_V})
-\mu_3\eta_3 Y([3];[1];c^q_{R_V},c^q_{L_V}),
$$
$$
{\Y}^{qV}_{31}=\sum_{\lambda=\pm}
Y([3];p_2,\lambda;1,1)
Y(p_2,\lambda;[1];c^q_{R_V},c^q_{L_V}),
$$
$$
{\tilde{\Y}}^{qV}_{31}={\Y}^{qV}_{31}
-\frac{{\cal F}^{qV}_{31}}{M_V^2}p_2\cdot(p_4+p_5),
$$
$$
{\Z}_{24}=Z([2];(2);\{1\};\{2\};1,1;1,1),
$$
$$
{\Z}_{312}^{qV}=Z([3];[1];
[2];(2);c^q_{R_V},c^q_{L_V};1,1),
$$
$$
{\tilde{\Z}}_{312}^{qV}={\Z}_{312}^{qV}-\frac{{\cal F}^{qV}_{31}}{M_V^2}
({\Y}_2+{\Y}_2'),
$$
$$
{\Z}_{314}^{qV}=Z([3];[1];
\{1\};\{2\};c^q_{R_V},c^q_{L_V};1,1),
$$
\be
{\tilde{\Z}}_{314}^{qV}={\Z}_{314}^{qV}-\frac{{\cal F}^{qV}_{31}}{M_V^2}
({\X}_4-{\Y}_4),
\ene
where $V$ represents a gauge boson $W^\pm$, $Z^0$ or $\gamma$, $q=u$ or
$d$ ($u$-- and $d$--type quarks of arbitrary masses $m_u$ and $m_d$,
respectively), and with the short--hand notations
$[x]=p_{x},\lambda_{x}$ $(x=1,...4)$, $(x)=q_x,\lambda_x$ $(x=1,2)$
and $\{x\}=r_{x},-$ $(x=1,2)$.\par
In the following we adopt $[i]=p_i,\lambda$ and $[j]=p_j,\lambda'$, whereas
the couplings $c_R$, $c_L$ and ${\cal H}$ can be easily
deduced from tabs.~VI--VII. Also, we sometimes make use of the equalities
\be
\Y_2+\Y_2'=\X_2,\quad\quad\quad
\X_{31}^{qV}+\X_{31}^{qV'}=\Y_{31}^{qV}+\F_{31}^{qV}.
\ene
\vskip1.0cm
\centerline{\sl 1. Process $d\gamma\rightarrow uW^-\phi$.}
\vskip 0.5cm
\noindent
In order to obtain from fig.~1 the Feynman graphs of  the process
\begin{equation}
d(p_1,\lambda_1) + \gamma (p_2,\lambda_2)
\longrightarrow u (p_3,\lambda_3) + W^- (p_4) + \phi (p_5),
\end{equation}
one has to make the following
assignments:
\be\label{ass:WH}
q=d,\quad\quad q'=u,\quad\quad V^{(*)}=W^{\pm(*)}.
\ene
The corresponding matrix element, summed over
 final spins and averaged over initial ones,
is given by
\begin{equation}\label{M2:dph->uWH}
{\left|{\overline M}\right|}=
{e^6\over 4}
N_2^2{3\over {8\pi M^2_{W^\pm}}}
\sum_{\{\lambda\}}
\int d\Omega_{r_1(r_2)}
\sum_{l,m=1}^{12}{T}_l^{\{\lambda\}} T_m^{\{\lambda\}*},
\end{equation}
where
$${\rm i}T_1^{\{\lambda\}}=
D_{u}(p_3+p_5)D_d(p_1+p_2)M_1^{\{\lambda\}}{\cal H}_1,\quad\quad
{\rm i}T_2^{\{\lambda\}}=
D_{d}(p_3+p_4)D_d(p_1+p_2)M_2^{\{\lambda\}}{\cal H}_2,$$
$${\rm i}T_3^{\{\lambda\}}=
D_{d}(p_3+p_4)D_d(p_1-p_5)M_3^{\{\lambda\}}{\cal H}_3,\quad\quad
{\rm i}T_4^{\{\lambda\}}=
D_{u}(p_3+p_5)D_u(p_1-p_4)M_4^{\{\lambda\}}{\cal H}_4,$$
$${\rm i}T_5^{\{\lambda\}}=
D_{u}(p_3-p_2)D_u(p_1-p_4)M_5^{\{\lambda\}}{\cal H}_5,\quad\quad
{\rm i}T_6^{\{\lambda\}}=
D_{u}(p_3-p_2)D_d(p_1-p_5)M_6^{\{\lambda\}}{\cal H}_6,$$
$${\rm i}T_7^{\{\lambda\}}=
D_{W^\pm}(p_4+p_5)D_d(p_1+p_2)M_7^{\{\lambda\}}{\cal H}_7,\quad\quad
{\rm i}T_8^{\{\lambda\}}=
D_{W^\pm}(p_4+p_5)D_u(p_3-p_2)M_8^{\{\lambda\}}{\cal H}_8,$$
$${\rm i}T_{9}^{\{\lambda\}}=
D_{W^\pm}(p_2-p_4)D_{u}(p_3+p_5)M_{9}^{\{\lambda\}}{\cal H}_{9},\quad\quad
{\rm i}T_{10}^{\{\lambda\}}=
D_{W^\pm}(p_2-p_4)D_{d}(p_1-p_5)M_{10}^{\{\lambda\}}{\cal H}_{10},$$
\begin{equation}
{\rm i}T_{11}^{\{\lambda\}}=
D_{W^\pm}(p_1-p_3)D_{W^\pm}(p_4+p_5)M_{11}^{\{\lambda\}}{\cal H}_{11},
\quad\quad
{\rm i}T_{12}^{\{\lambda\}}=
D_{W^\pm}(p_1-p_3)D_{W^\pm}(p_2-p_4)M_{12}^{\{\lambda\}}{\cal H}_{12}.
\end{equation}
We have
$$\hskip-2.0cm
M_{1}^{\{\lambda\}}=
\sum_{\lambda=\pm}\sum_{\lambda'=\pm}
\sum_{i=3,5,7}\sum_{j=1,2}(-b_ib_j)
Y([3];[i];c^{u}_{R_{\phi}},c^{u}_{L_{\phi}})
$$
$$\hskip1.5cm\times
Z([i];[j];
\{1\};\{2\};c_{R_{W^\pm}},c_{L_{W^\pm}};1,1)
Z([j];[1];
[2];(2);c^{d}_{R_{\gamma}},c^{d}_{L_{\gamma}};1,1),
$$
$$\hskip-1.0cm
M_{2}^{\{\lambda\}}=
\sum_{\lambda=\pm}\sum_{\lambda'=\pm}
\sum_{i=3,4,6}\sum_{j=1,2}(-b_ib_j)
Z([3];[i];
\{1\};\{2\};c_{R_{W^\pm}},c_{L_{W^\pm}};1,1)
$$
$$\hskip1.5cm\times
Y([i];[j];c^{d}_{R_{\phi}},c^{d}_{L_{\phi}})
Z([j];[1];
[2];(2);c^d_{R_{\gamma}},c^d_{L_{\gamma}};1,1),
$$
$$\hskip-1.0cm
M_{3}^{\{\lambda\}}=
\sum_{\lambda=\pm}\sum_{\lambda'=\pm}
\sum_{i=3,4,6}\sum_{j=1,5}(-b_ib_j)
Z([3];[i];
\{1\};\{2\};c_{R_{W^\pm}},c_{L_{W^\pm}};1,1)
$$
$$\hskip1.5cm\times
Z([i];[j];
[2];(2);c^d_{R_{\gamma}},c^d_{L_{\gamma}};1,1)
Y([j];[1];c^d_{R_{\phi}},c^d_{L_{\phi}}),
$$
$$\hskip-2.0cm
M_{4}^{\{\lambda\}}=
\sum_{\lambda=\pm}\sum_{\lambda'=\pm}
\sum_{i=3,5,7}\sum_{j=1,4,6}(-b_ib_j)
Y([3];[i];c^u_{R_{\phi}},c^u_{L_{\phi}})
$$
$$\hskip1.5cm\times
Z([i];[j];
[2];(2);c^u_{R_{\gamma}},c^u_{L_{\gamma}};1,1)
Z([j];[1];
\{1\};\{2\};c_{R_{W^\pm}},c_{L_{W^\pm}};1,1),
$$
$$\hskip-2.0cm
M_{5}^{\{\lambda\}}=
\sum_{\lambda=\pm}\sum_{\lambda'=\pm}
\sum_{i=3,2}\sum_{j=1,4,6}(-b_ib_j)
Z([3];[i];
[2];(2);c^u_{R_{\gamma}},c^u_{L_{\gamma}};1,1)
$$
$$\hskip1.5cm\times
Y([i];[j];c^u_{R_{\phi}},c^u_{L_{\phi}})
Z([j];[1];
\{1\};\{2\};c_{R_{W^\pm}},c_{L_{W^\pm}};1,1),
$$
$$\hskip-2.0cm
M_{6}^{\{\lambda\}}=
\sum_{\lambda=\pm}\sum_{\lambda'=\pm}
\sum_{i=3,2}\sum_{j=1,5,7}(-b_ib_j)
Z([3];[i];
[2];(2);c^u_{R_{\gamma}},c^u_{L_{\gamma}};1,1)
$$
$$\hskip1.5cm\times
Z([i];[j];
\{1\};\{2\};c_{R_{W^\pm}},c_{L_{W^\pm}};1,1)
Y([j];[1];c^d_{R_{\phi}},c^d_{L_{\phi}}),
$$
$$\hskip-2.5cm
M_7^{\{\lambda\}}=\sum_{\lambda=\pm}\sum_{i=1,2}
(-b_i)Z([i];[1];[2];(2);c^d_{R_\gamma},c^d_{L_\gamma};1,1)
$$
$$
\times
\{Z([3];[i];\{1\};\{2\};c_{R_{W^\pm}},c_{L_{W^\pm}};1,1)
$$
$$\hskip1.5cm
-{{{\X}_4}\over{M_{W^\pm}^2}}
[\sum_{\lambda'=\pm}\sum_{j=1,2,3}(-b_j)
Y([3];[j];1,1)Y([j];[i];c_{R_{W^\pm}},c_{L_{W^\pm}})]\},
$$
$$\hskip-2.5cm
M_8^{\{\lambda\}}=\sum_{\lambda=\pm}\sum_{i=2,3}
(b_i)Z([3];[i];[2];(2);c^u_{R_\gamma},c^u_{L_\gamma};1,1)
$$
$$
\times
\{Z([i];[1];\{1\};\{2\};c_{R_{W^\pm}},c_{L_{W^\pm}};1,1)
$$
$$
\hskip1.5cm
-{{{\X}_4}\over{M_{W^\pm}^2}}
[\sum_{\lambda'=\pm}\sum_{j=1,2,3}(-b_j)
Y([i];[j];1,1)Y([j];[1];c_{R_{W^\pm}},c_{L_{W^\pm}})]\},
$$
$$\hskip-2.5cm
M_{9}^{\{\lambda\}}=\sum_{\lambda=\pm}\sum_{i=3,5,7}(2b_i)
Y([3];[i];c^u_{R_{\phi}},c^u_{L_{\phi}})
$$
$$\times
[\Z_{24}\sum_{\lambda'=\pm}
Y([i];p_2,\lambda';1,1)Y(p_2,\lambda';[1];c_{R_{W^\pm}},c_{L_{W^\pm}})
$$
$$\hskip.5cm
-\Y_2Z(\{1\};\{2\};[i];[1];1,1;c_{R_{W^\pm}},c_{L_{W^\pm}})
-\Y_4Z([2];(2);[i];[1];1,1;c_{R_{W^\pm}},c_{L_{W^\pm}})],
$$
$$\hskip-2.5cm
M_{10}^{\{\lambda\}}=\sum_{\lambda=\pm}\sum_{i=1,5,7}(-2b_i)
Y([i];[1];c^d_{R_{\phi}},c^d_{L_{\phi}})
$$
$$\times
[\Z_{24}\sum_{\lambda'=\pm}
Y([3];p_2,\lambda';1,1)Y(p_2,\lambda';[i];c_{R_{W^\pm}},c_{L_{W^\pm}})
$$
$$\hskip.5cm
-\Y_2Z(\{1\};\{2\};[3];[i];1,1;c_{R_{W^\pm}},c_{L_{W^\pm}})
-\Y_4Z([2];(2);[3];[i];1,1;c_{R_{W^\pm}},c_{L_{W^\pm}})],
$$
$$\hskip-2.5cm
M_{11}^{\{\lambda\}}=\Z_{24}(\F_{31}^{W^{\pm}}+2\Y_{31}^{W^{\pm}})
-2\X_2\tilde{\Z}_{314}^{W^{\pm}}-(2\Y_4-\X_4)\tilde{\Z}_{312}^{W^{\pm}}
$$
$$
-{1\over M_{W^\pm}^2}
[\X_2\X_4(\Y_{31}^{W^{\pm}}-\X_{31}^{W^{\pm}}-\X_{31}^{W^{\pm{'}}})
+(p_1-p_3)^2(\Z_{24}\F_{31}^{W^{\pm}}+\tilde{\Z}_{312}^{W^{\pm}}\X_4)
+2p_2\cdot(p_1-p_3)\Z_{24}\F_{31}^{W^{\pm}}]
$$
$$\hskip1.5cm
-{1\over M_{W^\pm}^4}
\{[(p_1-p_3)^2+p_2\cdot(p_1-p_3)]
\X_4(\Y_2-\Y_2')\F_{31}^{W^{\pm}}\},
$$
\be\label{M:dph->uWH}
M_{12}^{\{\lambda\}}=2({\tilde{\Y}}^{W^{\pm}}_{31}{\Z}_{24}-
{\Y}_{2}{\tilde{\Z}}^{W^{\pm}}_{314}-{\Y}_{4}{\tilde{\Z}}^{W^{\pm}}_{312}).
\ene

\vskip 1.0cm
\centerline{\sl 2. Process $d\gamma\rightarrow dZ^0\phi$.}
\vskip 0.5cm
\noindent
The Feynman graphs for  the process
\begin{equation}
d(p_1,\lambda_1) + \gamma (p_2,\lambda_2)
\longrightarrow d (p_3,\lambda_3) + Z^0 (p_4) + \phi (p_5),
\end{equation}
can be obtained from fig.~1 by setting
\be
q=q'=d,\quad\quad\quad V^{(*)}=Z^{0(*)}.
\ene
The formulae for the amplitude squared are practically the same as in
the previous section,  considering the first 8
amplitudes only (i.e., $M_{i}^{\{\lambda\}}=0$ for $i=9,...12$),
with the relabeling:
\be\label{ch:dph->dZH}
u\ar d,\quad\quad W^\pm\ar Z^0,
\ene
in eqs.~(\ref{M2:dph->uWH})--(\ref{M:dph->uWH}).
\vskip 1.0cm
\centerline{\sl 3. Process $g\gamma\rightarrow u\bar u\phi$.}
\vskip 0.5cm
The Feynman diagrams for
\begin{equation}
g(p_1,\lambda_1) + \gamma (p_2,\lambda_2)
\longrightarrow u (p_3,\lambda_3) + \bar u (p_4,\lambda_4) + \phi (p_5),
\end{equation}
are shown in fig.~2, where $q=u$.
The amplitude squared is
\begin{equation}\label{M2:gph->udH}
{\left|{\overline M}\right|}=
{e^4g_s^2\over 4}
N_1^2N_2^2
\sum_{\{\lambda\}}
\sum_{l,m=1}^{6}{T}_l^{\{\lambda\}} T_m^{\{\lambda\}*}.
\end{equation}
The expressions for the $T_i^{\{\lambda\}}$ 's are
$${\rm -i}T_1^{\{\lambda\}}=
D_u(p_3+p_5)D_u(p_1-p_4)M_{1}^{\{\lambda\}}{\cal H}_1,\quad\quad
{\rm -i}T_2^{\{\lambda\}}=
D_u(p_3-p_2)D_u(p_1-p_4)M_{2}^{\{\lambda\}}{\cal H}_2,$$
$${\rm -i}T_3^{\{\lambda\}}=
D_u(p_3-p_2)D_u(p_4+p_5)M_{3}^{\{\lambda\}}{\cal H}_3,$$
\be
{\rm -i}T_{i+3}^{\{\lambda\}}={\rm -i}T_i^{\{\lambda\}}
(p_3\leftrightarrow p_4),
\quad\quad\quad i=1,...3,
\ene
while the spinor functions are
$$\hskip-2.0cm
M_{1}^{\{\lambda\}}=
\sum_{\lambda=\pm}\sum_{\lambda'=\pm}
\sum_{i=3,5,7}\sum_{j=1,4}(-b_ib_j)
Y([3];[i];c^u_{R_{\phi}},c^u_{L_{\phi}})
$$
$$\hskip1.5cm\times
Z([i];[j];[2];(2);c^{u}_{R_{\gamma}},c^{u}_{L_{\gamma}};1,1)
Z([j];[4];[1];(1);c^{u}_{R_{g}},c^{u}_{L_{g}};1,1),
$$
$$\hskip-2.0cm
M_{2}^{\{\lambda\}}=
\sum_{\lambda=\pm}\sum_{\lambda'=\pm}
\sum_{i=2,3}\sum_{j=1,4}(-b_ib_j)
Z([3];[i];[2];(2);c^{u}_{R_{\gamma}},c^{u}_{L_{\gamma}};1,1)
$$
$$\hskip1.5cm\times
Y([i];[j];c^u_{R_{\phi}},c^u_{L_{\phi}})
Z([j];[4];[1];(1);c^{u}_{R_{g}},c^{u}_{L_{g}};1,1),
$$
$$\hskip-2.0cm
M_{3}^{\{\lambda\}}=
\sum_{\lambda=\pm}\sum_{\lambda'=\pm}
\sum_{i=2,3}\sum_{j=4,5,7}(-b_ib_j)
Z([3];[i];[2];(2);c^{u}_{R_{\gamma}},c^{u}_{L_{\gamma}};1,1)
$$
$$\hskip1.5cm\times
Z([i];[j];[1];(1);c^{u}_{R_{g}},c^{u}_{L_{g}};1,1)
Y([j];[4];c^u_{R_{\phi}},c^u_{L_{\phi}}),
$$
\be
M_{i+3}^{\{\lambda\}}=M_{i}^{\{\lambda\}}
(p_3\leftrightarrow p_4),
\quad\quad\quad i=1,...3.
\ene
\vskip1.0cm
\noindent
By trivial relabeling and sign exchanges, it is possible
to obtain from the above formulae the corresponding
ones for the $u$--type quark initiated processes
$$u\gamma\rightarrow dW^+\phi,$$
\be
u\gamma\rightarrow uZ^0\phi,
\ene
as for the charge conjugate reactions
$$\bar d\gamma\rightarrow \bar uW^+\phi,$$
\be
\bar d\gamma\rightarrow \bar dZ^0\phi.
\ene
Finally, the same it can be done for obtaining the helicity amplitudes
for the $g$--initiated process
\be
g\gamma\rightarrow d\bar d \phi.
\ene

\vfill
\newpage
\subsection*{Table Captions}
\begin{description}
\item[Tab.~I  ] Production cross sections for processes
(\ref{proc1})--(\ref{proc3}), at $\sqrt s_{ep}=1.36$ TeV, with $M_\phi=60,
80,100,120$ and 140 GeV. The HMRS(B) structure functions are used.
The errors are the statistical errors on the numerical calculation.
\item[Tab.~II ] Production cross sections for processes
(\ref{proc1})--(\ref{proc3}) in (a)--(c), respectively,
at $\sqrt s_{ep}=1.36$ TeV, with $M_\phi=60$ GeV, for all different
flavour combinations entering in the partonic subprocesses.
The HMRS(B) structure functions are used.
The errors are the statistical errors on the numerical calculation.
\item[Tab.~III] Production cross sections for the background processes
discussed in the text. The HMRS(B) structure functions are used.
The errors are the statistical errors on the numerical calculation.
\item[Tab.~IV ] Number of signal ($S$) and background events ($B$)
and their statistical significance ($S/\sqrt B$), for the processes
(\ref{proc1})--(\ref{proc3}), at $\sqrt s_{ep}=1.36$ TeV, in the window
$|M_{b\bar b}-M_\phi|<5$ GeV, for the usual selection
of Higgs masses. Numbers correspond to
hadronic(leptonic) decays of the $W^\pm/Z^0$'s. The HMRS(B) structure
functions are used.
The symbol ``--'' indicates the case in which the backgrounds do not
constitute a problem in disentangling the signals.
\item[Tab.~V  ] Total number of signal ($S_{\rm tot}$) and background events
($B_{\rm tot}$) and their statistical significance ($S_{\rm tot}/
\sqrt B_{\rm tot}$), after summing the numbers in tab.~IV in ``inclusive''
rates.
\item[Tab.~VI ] \sm\ Higgs boson ${\cal H}$--couplings
to the gauge bosons $W^\pm$ and $Z^0$.
\item[Tab.~VII] \sm\ right and left handed couplings $(c_R,c_L)$
of $u$-- and  $d$--type  quarks to the neutral
gauge bosons $g$, $\gamma$, $Z^0$, to the charged $W^\pm$'s and to
the Higgs boson $\phi$.
We have $g_R^q=-Q^qs^2_W$ and $g_L^q=T^q_3-Q^qs^2_W$ ($q=u,d$), with
$(Q^u, T^u)=(+{2\over 3}, {1\over 2})$ and
$(Q^d, T^d_3)=(-{1\over 3}, -{1\over 2})$ for quark
charges and isospins.
\end{description}
\vfill
\newpage
\subsection*{Figure Captions}
\begin{description}
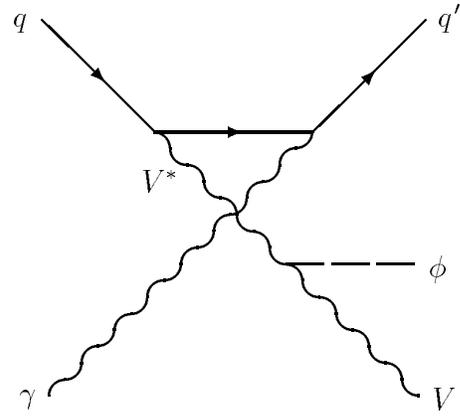
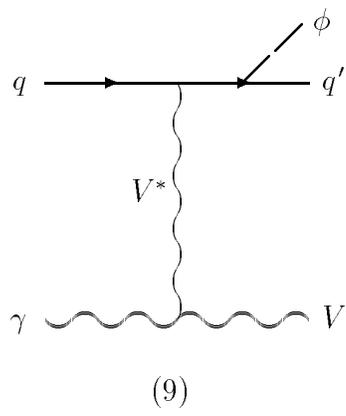
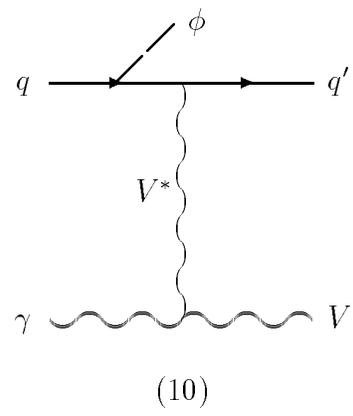
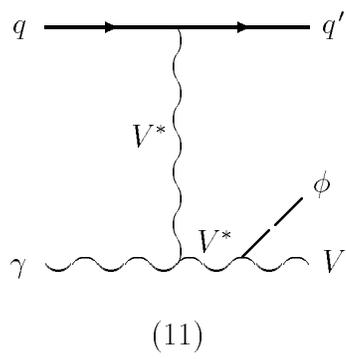
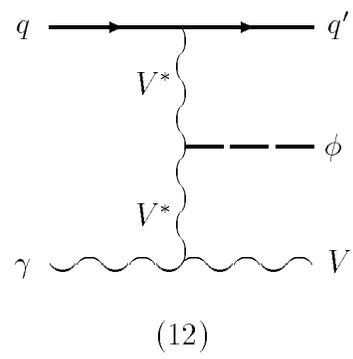
\item[Fig.~1 ] Feynman diagrams contributing in lowest order to
$q\gamma\rightarrow q'V\phi$, where $q(q')$ represents a quark,
$V(V^*)$ an external(internal) vector boson and $\phi$ the
\sm\ Higgs boson, in the unitary gauge.
In the case $V=Z^0$ and $q'=q$ only the
first eight diagrams of fig.~1 contribute.
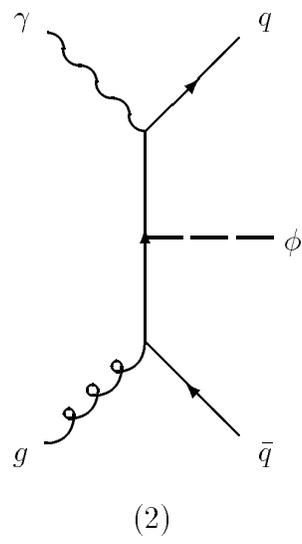
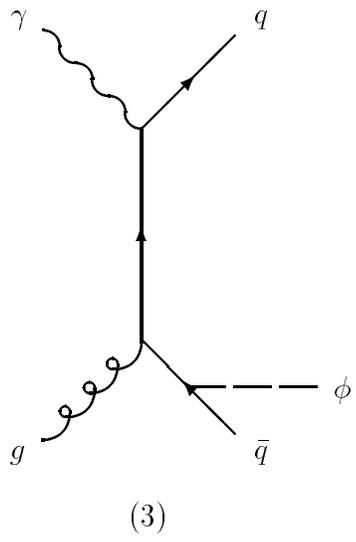
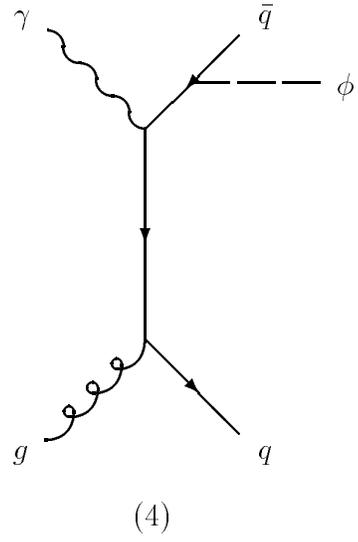
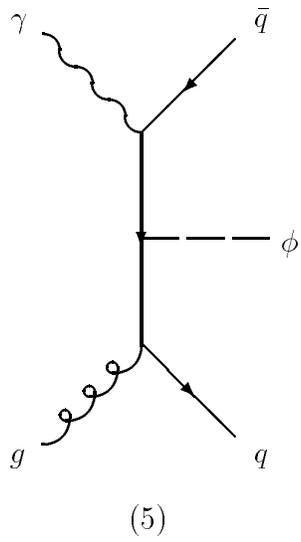
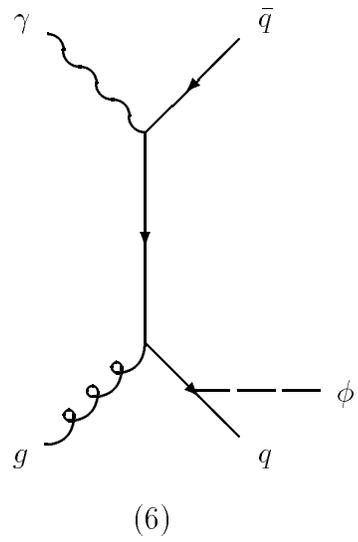
\item[Fig.~2 ] Feynman diagrams contributing in the lowest order to
$g\gamma\rightarrow q\bar q\phi$, where $q$ represents a quark and
$\phi$ the \sm\ Higgs boson, in the unitary gauge.
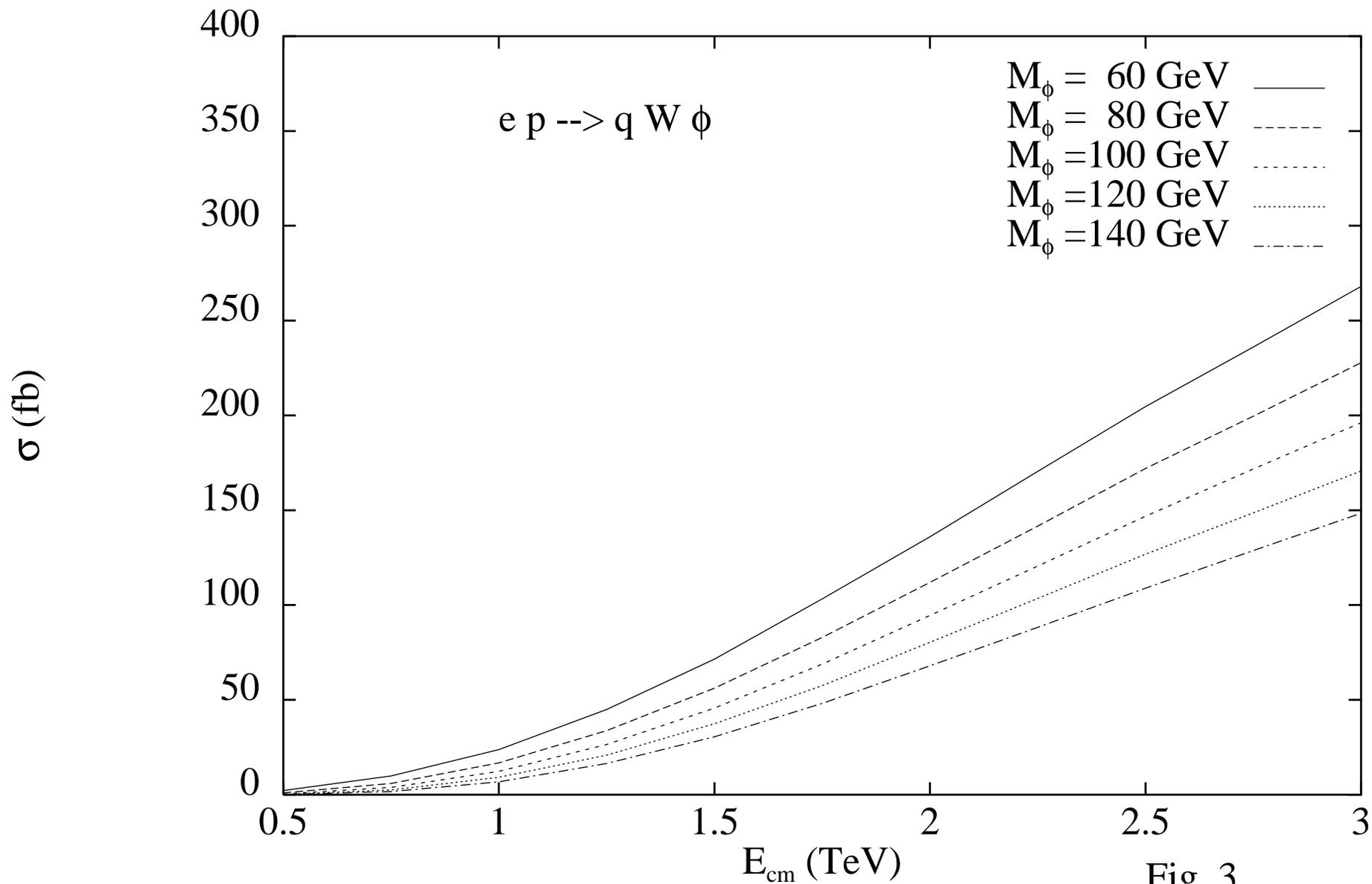
\item[Fig.~3 ] Cross sections of process (\ref{proc1}) as a function
of $\sqrt s_{ep}$, for a selection of Higgs masses.
The HMRS(B) structure functions are used.
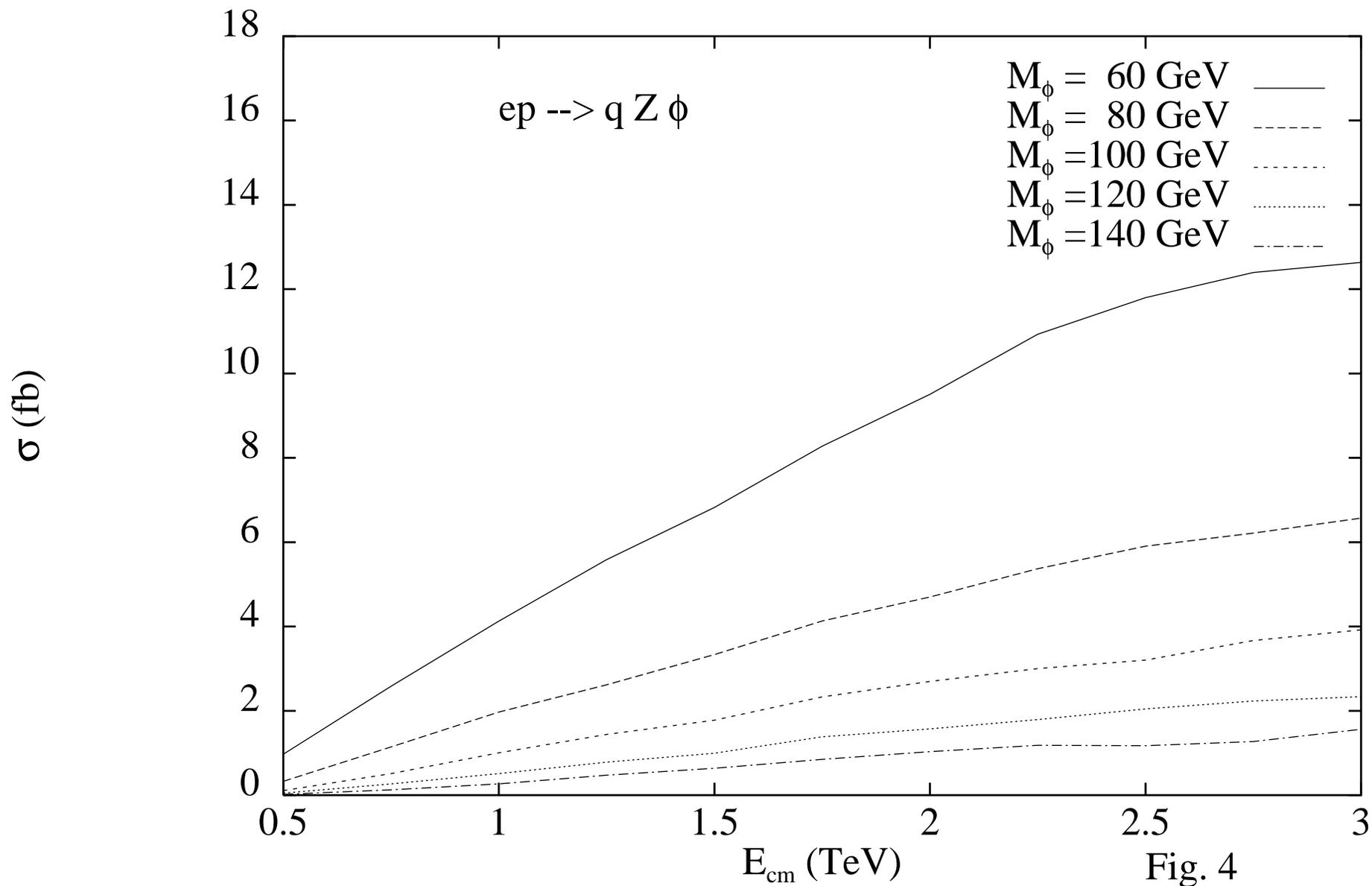
\item[Fig.~4 ] Cross sections of process (\ref{proc2}) as a function
of $\sqrt s_{ep}$, for a selection of Higgs masses.
The HMRS(B) structure functions are used.
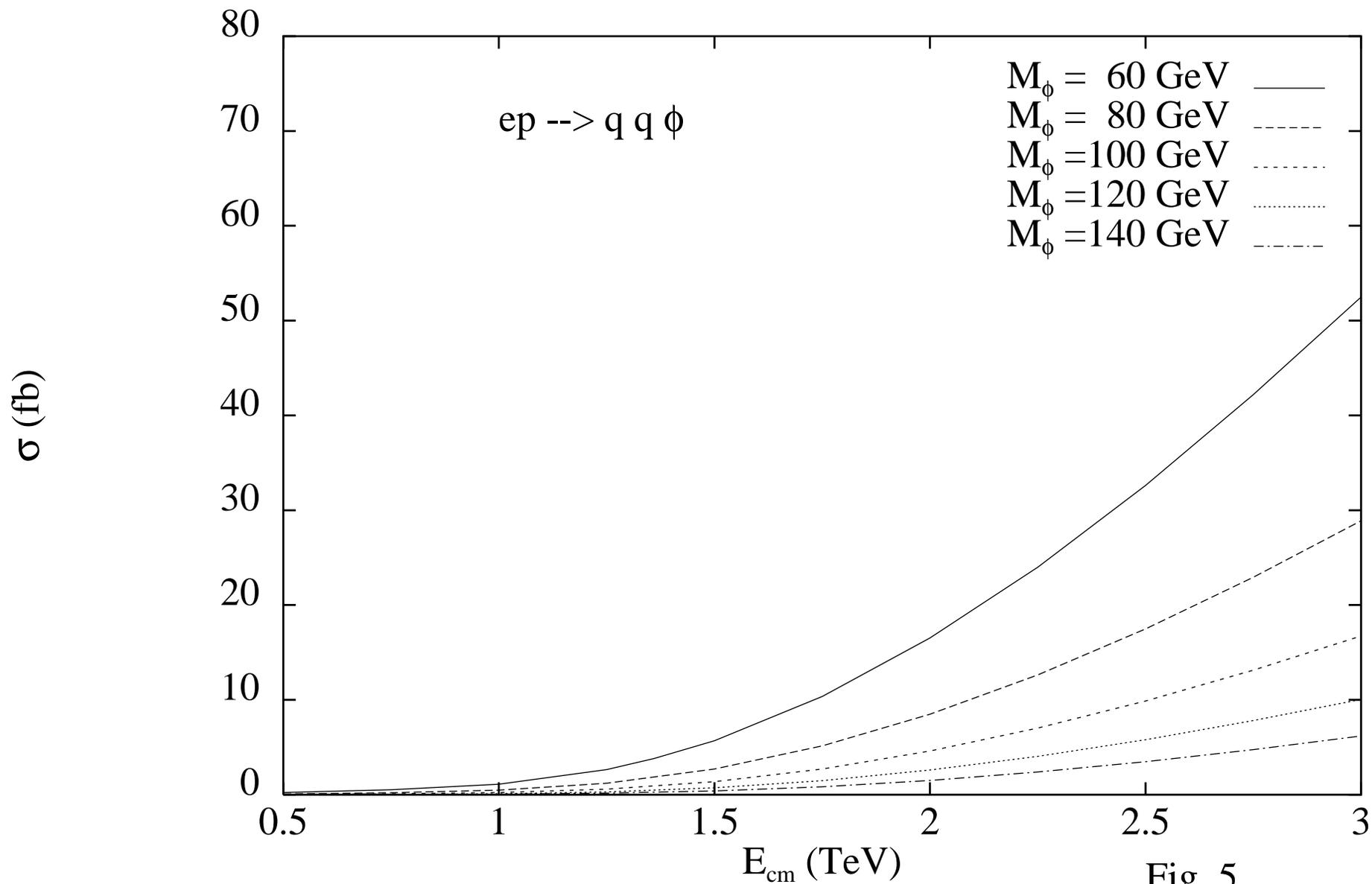
\item[Fig.~5 ] Cross sections of process (\ref{proc3}) as a function
of $\sqrt s_{ep}$, for a selection of Higgs masses.
The HMRS(B) structure functions are used.
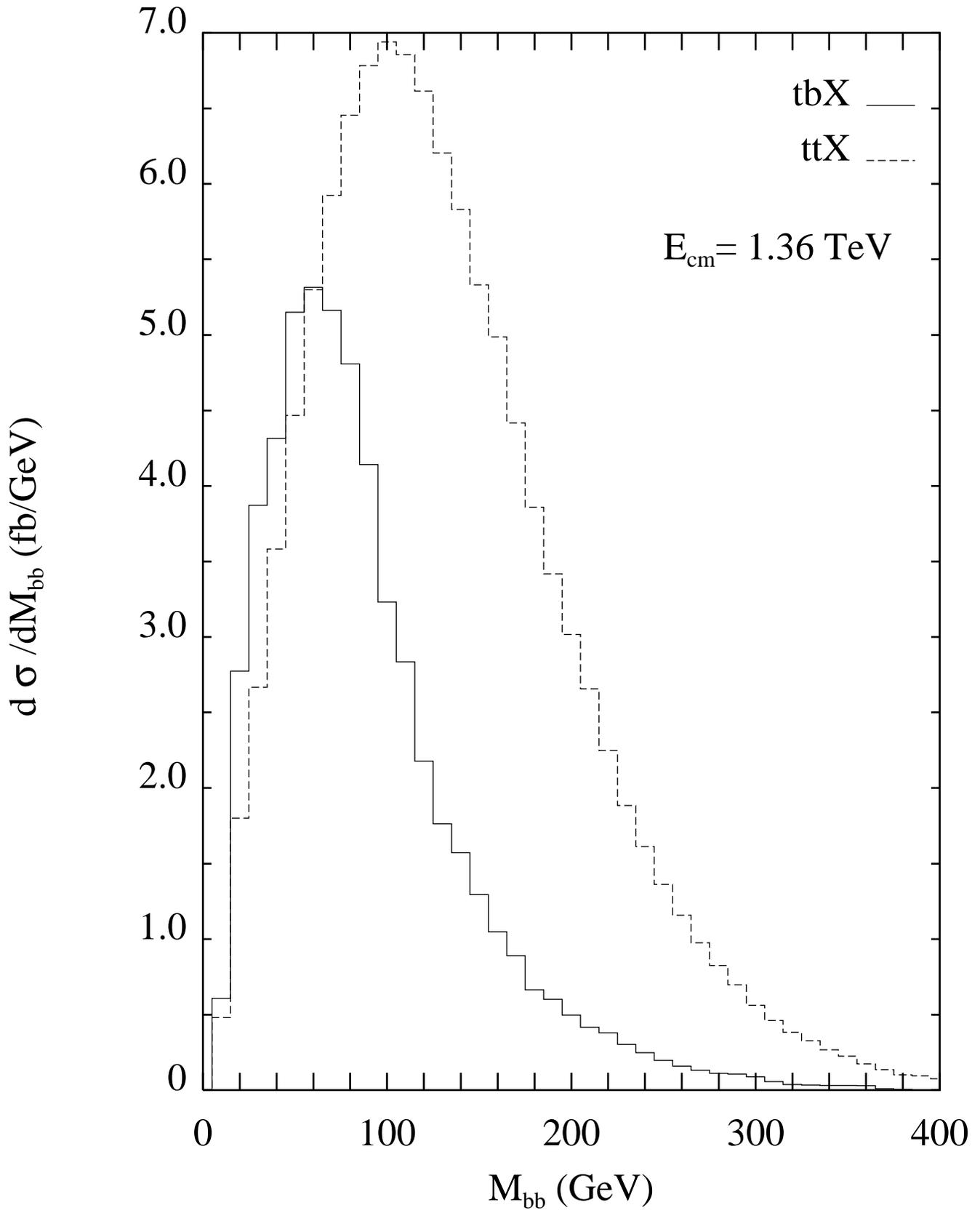
\item[Fig.~6 ] Differential distributions in the invariant mass of the
$b\bar b$--pair $M_{b\bar b}$ for the $\bar tbX\ar b\bar b W^\pm X$ and
$t\bar tX\ar b\bar bW^\pm X$ backgrounds,
at $\sqrt s_{ep}=1.36$ TeV.
The HMRS(B) structure functions are used.
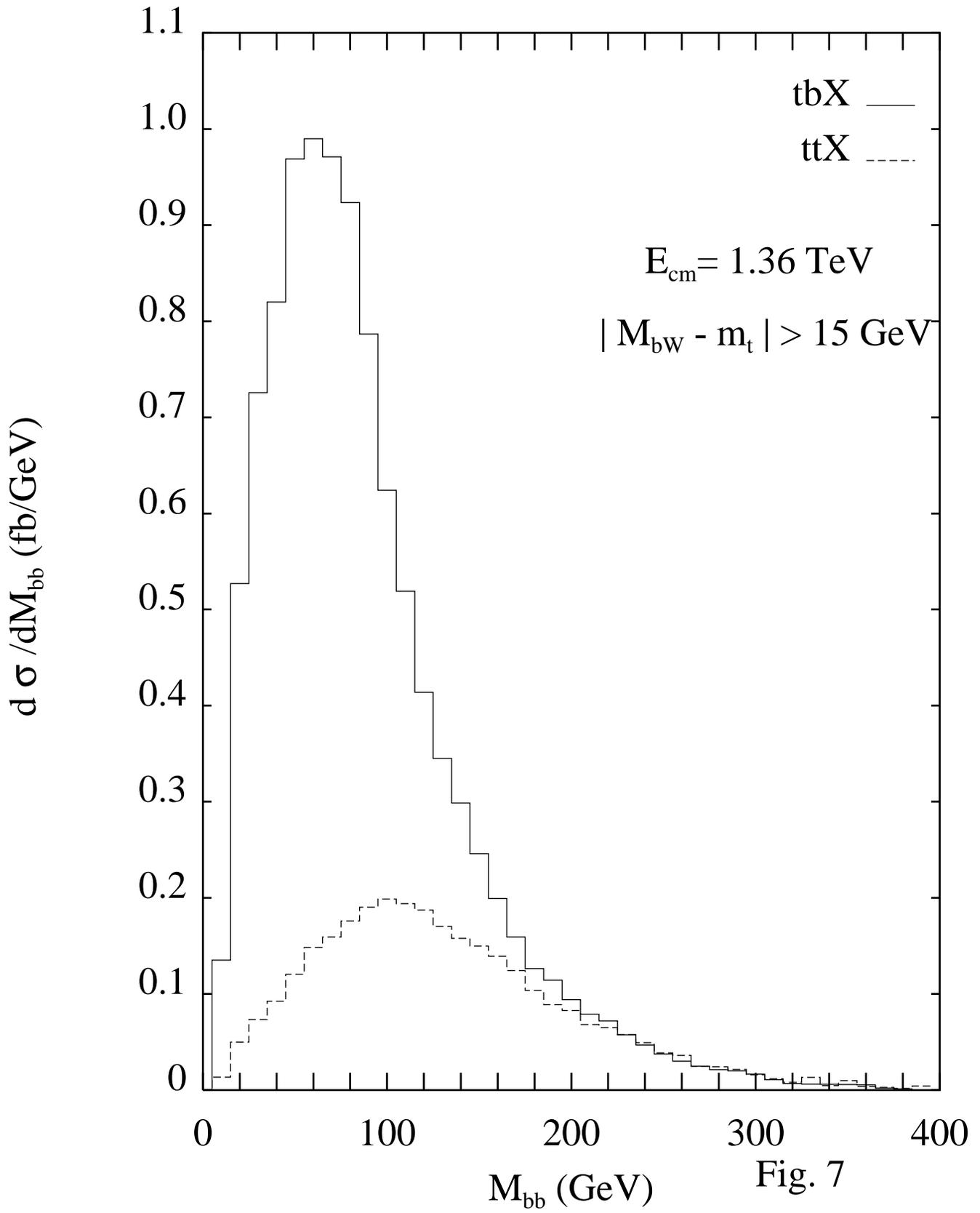
\item[Fig.~7 ] Differential distributions in the invariant mass of the
$b\bar b$--pair $M_{b\bar b}$ for the $\bar tbX\ar b\bar b W^\pm X$ and
$t\bar tX\ar b\bar bW^\pm $ backgrounds, at $\sqrt s_{ep}=1.36$ TeV,
after the cut $|M_{bW\ar bjj}-m_t|>15$ GeV.
The HMRS(B) structure functions are used.
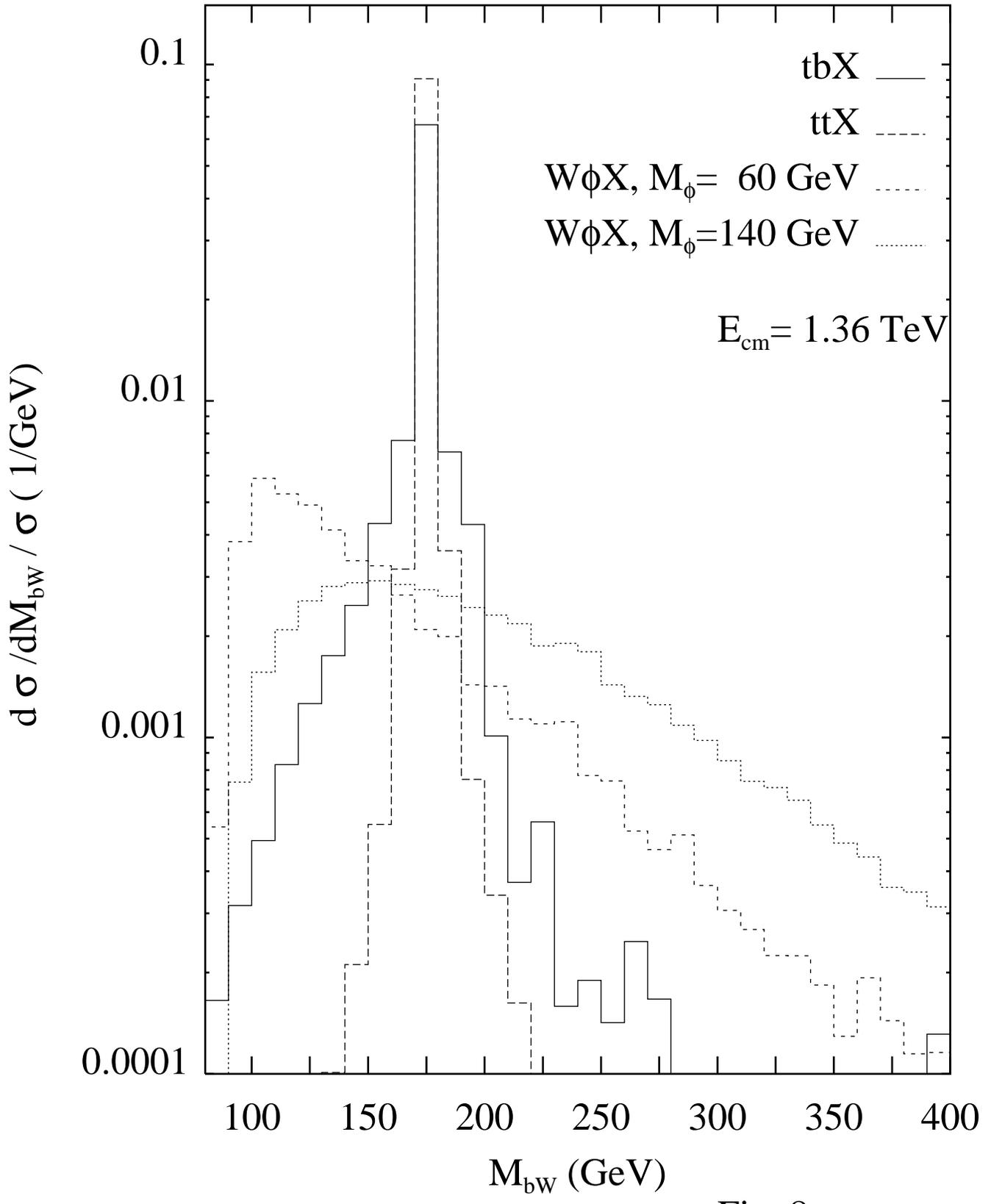
\item[Fig.~8 ] Differential distributions in the invariant mass of the
$b W$--system $M_{W b}$ for the $\bar tbX\ar b\bar b W^\pm X$ and
$t\bar tX\ar b\bar bW^\pm X$ backgrounds, and the signal
$W^\pm\phi X\ar W^\pm(b\bar b) X$ with $M_\phi=60, 140$ GeV,
at $\sqrt s_{ep}=1.36$ TeV.
The HMRS(B) structure functions are used.
\end{description}

\vspace*{\fill}

\vfill
\newpage
\thispagestyle{empty}

\
\vskip4.0cm
\begin{table}%[p]%[htbp]
\begin{center}
\begin{tabular}{|c|c|c|c|}
\hline
\multicolumn{4}{|c|}
{\rule[-0.5cm]{0cm}{1.3cm}
$\sigma$ (fb)}
\\ \hline
\rule[-0.6cm]{0cm}{1.3cm}
$M_{\phi}$ (GeV) & $q'W^\pm\phi$ &  $q Z^0\phi$ &  $q\bar q \phi$ \\ \hline
\rule[-0.6cm]{0cm}{1.3cm}
$60$  & $55.61\pm0.34$ & $6.13\pm0.10$ & $3.806\pm0.058$  \\ \hline
\rule[-0.6cm]{0cm}{1.3cm}
$80$  & $42.84\pm0.25$ & $3.056\pm0.052$ & $1.765\pm0.029$  \\ \hline
\rule[-0.6cm]{0cm}{1.3cm}
$100$ & $34.53\pm0.14$ & $1.581\pm0.028$ & $0.872\pm0.013$  \\ \hline
\rule[-0.6cm]{0cm}{1.3cm}
$120$ & $27.56\pm0.11$ & $0.798\pm0.024$ & $0.4513\pm0.0068$  \\ \hline
\rule[-0.6cm]{0cm}{1.3cm}
$140$ & $22.048\pm0.080$ & $0.547\pm0.018$ & $0.2419\pm0.0039$  \\ \hline
\multicolumn{4}{|c|}
{\rule[-0.5cm]{0cm}{1.3cm}
$\sqrt s=1.36$ TeV\quad\quad\quad\quad HMRS(B)}
 \\ \hline
\multicolumn{4}{c}
{\rule{0cm}{.9cm}
{\Large Table I}}  \\
\multicolumn{4}{c}
{\rule{0cm}{.9cm}}

\end{tabular}
\end{center}
\end{table}

\vfill
\newpage
\thispagestyle{empty}

\
\vskip4.0cm
\begin{table}%[p]%[htbp]
\begin{center}
\begin{tabular}{|c|c|}
\hline
\rule[-0.6cm]{0cm}{1.3cm}
{\rm Flavours} &  $\sigma$ (fb) \\ \hline
\rule[-0.6cm]{0cm}{1.3cm}
$u\gamma \ar dW^+\phi + \bar u\gamma \ar \bar dW^-\phi$
& $29.58\pm0.15$  \\ \hline
\rule[-0.6cm]{0cm}{1.3cm}
$d\gamma \ar uW^-\phi + \bar d\gamma \ar \bar uW^+\phi$
& $19.37\pm0.30$   \\ \hline
\rule[-0.6cm]{0cm}{1.3cm}
$s\gamma \ar cW^-\phi + \bar s\gamma \ar \bar cW^+\phi$
& $4.228\pm0.021$  \\ \hline
\rule[-0.6cm]{0cm}{1.3cm}
$c\gamma \ar sW^+\phi + \bar c\gamma \ar \bar sW^-\phi$
& $1.620\pm0.012$  \\ \hline
\rule[-0.6cm]{0cm}{1.3cm}
$b\gamma \ar tW^-\phi + \bar b\gamma \ar \bar tW^+\phi$
& $0.7995\pm0.0033$ \\ \hline
\multicolumn{2}{|c|}
{\rule[-0.5cm]{0cm}{1.3cm}
$\sqrt s=1.36$ TeV\quad\quad HMRS(B)\quad\quad $M_{\phi}=60$ GeV}
 \\ \hline
\multicolumn{2}{c}
{\rule{0cm}{.9cm}
{\Large Table IIa}}  \\
\multicolumn{2}{c}
{\rule{0cm}{.9cm}}

\end{tabular}
\end{center}
\end{table}

\vfill
\newpage
\thispagestyle{empty}

\
\vskip4.0cm
\begin{table}%[p]%[htbp]
\begin{center}
\begin{tabular}{|c|c|}
\hline
\rule[-0.6cm]{0cm}{1.3cm}
{\rm Flavours} &  $\sigma$ (fb) \\ \hline
\rule[-0.6cm]{0cm}{1.3cm}
$u\gamma \ar uZ^0\phi + \bar u\gamma \ar \bar uZ^0\phi$
& $4.535\pm0.097$  \\ \hline
\rule[-0.6cm]{0cm}{1.3cm}
$d\gamma \ar dZ^0\phi + \bar d\gamma \ar \bar dZ^0\phi$
& $0.982\pm0.025$   \\ \hline
\rule[-0.6cm]{0cm}{1.3cm}
$s\gamma \ar sZ^0\phi + \bar s\gamma \ar \bar sZ^0\phi$
& $0.2707\pm0.0015$  \\ \hline
\rule[-0.6cm]{0cm}{1.3cm}
$c\gamma \ar cZ^0\phi + \bar c\gamma \ar \bar cZ^0\phi$
& $0.3018\pm0.0012$  \\ \hline
\rule[-0.6cm]{0cm}{1.3cm}
$b\gamma \ar bZ^0\phi + \bar b\gamma \ar \bar bZ^0\phi$
& $0.03839\pm0.00017$ \\ \hline
\multicolumn{2}{|c|}
{\rule[-0.5cm]{0cm}{1.3cm}
$\sqrt s=1.36$ TeV\quad\quad HMRS(B)\quad\quad $M_{\phi}=60$ GeV}
 \\ \hline
\multicolumn{2}{c}
{\rule{0cm}{.9cm}
{\Large Table IIb}}  \\
\multicolumn{2}{c}
{\rule{0cm}{.9cm}}

\end{tabular}
\end{center}
\end{table}

\vfill
\newpage
\thispagestyle{empty}

\
\vskip4.0cm
\begin{table}%[p]%[htbp]
\begin{center}
\begin{tabular}{|c|c|}
\hline
\rule[-0.6cm]{0cm}{1.3cm}
{\rm Flavours} &  $\sigma$ (fb) \\ \hline
\rule[-0.6cm]{0cm}{1.3cm}
$g\gamma \ar u\bar u\phi$ & $(60.4\pm2.2)\times10^{-6}$  \\ \hline
\rule[-0.6cm]{0cm}{1.3cm}
$g\gamma \ar d\bar d\phi$ & $(51.09\pm0.83)\times10^{-6}$   \\ \hline
\rule[-0.6cm]{0cm}{1.3cm}
$g\gamma \ar s\bar s\phi$ & $(11.113\pm0.071)\times10^{-3}$  \\ \hline
\rule[-0.6cm]{0cm}{1.3cm}
$g\gamma \ar c\bar c\phi$ & $0.6572\pm0.0025$  \\ \hline
\rule[-0.6cm]{0cm}{1.3cm}
$g\gamma \ar b\bar b\phi$ & $0.5188\pm0.0019$ \\ \hline
\rule[-0.6cm]{0cm}{1.3cm}
$g\gamma \ar t\bar t\phi$ & $2.6192\pm0.0049$ \\ \hline
\multicolumn{2}{|c|}
{\rule[-0.5cm]{0cm}{1.3cm}
$\sqrt s=1.36$ TeV\quad\quad HMRS(B)\quad\quad $M_{\phi}=60$ GeV}
 \\ \hline
\multicolumn{2}{c}
{\rule{0cm}{.9cm}
{\Large Table IIc}}  \\
\multicolumn{2}{c}
{\rule{0cm}{.9cm}}

\end{tabular}
\end{center}
\end{table}

\vfill
\newpage
\thispagestyle{empty}

\
\vskip4.0cm
\begin{table}%[p]%[htbp]
\begin{center}
\begin{tabular}{|c|c|}
\hline
\rule[-0.6cm]{0cm}{1.3cm}
{Background} &  $\sigma$ (fb) \\ \hline
\rule[-0.6cm]{0cm}{1.3cm}
$ep\ar W^\pm Z^0 X$ & $224.3\pm1.9$  \\ \hline
\rule[-0.6cm]{0cm}{1.3cm}
$ep\ar \bar tbX\ar b\bar b W^\pm X$ & $535.3\pm5.1$   \\ \hline
\rule[-0.6cm]{0cm}{1.3cm}
$ep\ar t\bar t X\ar b\bar b W^\pm X$   & $1114.7\pm1.4$  \\ \hline
\rule[-0.6cm]{0cm}{1.3cm}
$ep \ar Z^0 Z^0 X$  & $12.15\pm0.50$  \\ \hline
\rule[-0.6cm]{0cm}{1.3cm}
$ep\ar q\bar q Z^0X$ & $3714\pm91$ \\ \hline
\multicolumn{2}{|c|}
{\rule[-0.5cm]{0cm}{1.3cm}
$\sqrt s=1.36$ TeV\quad\quad HMRS(B)}
 \\ \hline
\multicolumn{2}{c}
{\rule{0cm}{.9cm}
{\Large Table III}}  \\
\multicolumn{2}{c}
{\rule{0cm}{.9cm}}

\end{tabular}
\end{center}
\end{table}

\newpage
\thispagestyle{empty}

\vskip0.01cm
\begin{table}%[p]%[htbp]
\begin{center}
\begin{tabular}{|c|c|c|c|c|}
\hline
\rule[-0.6cm]{0cm}{1.3cm}
{\rm Process} &  $S$  &  $B$  &  $S/\sqrt B$ &  $M_\phi$ (GeV) \\ \hline\hline
\rule[-0.6cm]{0cm}{1.3cm}
$q'W^\pm\phi$  &  $99(42)$  &  $418(179)$  &  $4.84(3.14)$ &  \\ \hline
\rule[-0.6cm]{0cm}{1.3cm}
$q Z^0\phi$    &  $11(2)$  &  $0(0)$  &  $-(-)$ &  $60$ \\ \hline
\rule[-0.6cm]{0cm}{1.3cm}
$q \bar q\phi$ &  $10$  &  $0$  &  $-$ &  \\ \hline\hline
\rule[-0.6cm]{0cm}{1.3cm}
$q'W^\pm\phi$  &  $75(32)$  &  $452(194)$  &  $3.53(2.30)$ &  \\ \hline
\rule[-0.6cm]{0cm}{1.3cm}
$q Z^0\phi$    &  $5(1)$  &  $0(0)$  &  $-(-)$ &  $80$ \\ \hline
\rule[-0.6cm]{0cm}{1.3cm}
$q \bar q\phi$ &  $4$  &  $0$  &  $-$ &  \\ \hline\hline
\rule[-0.6cm]{0cm}{1.3cm}
$q'W^\pm\phi$  &  $59(25)$  &  $412(177)$  &  $2.91(1.88)$ &  \\ \hline
\rule[-0.6cm]{0cm}{1.3cm}
$q Z^0\phi$    &  $3(0)$  &  $0(0)$  &  $-(0)$ &  $100$ \\ \hline
\rule[-0.6cm]{0cm}{1.3cm}
$q \bar q\phi$ &  $2$  &  $196$  &  $0.14$ &  \\ \hline\hline
\rule[-0.6cm]{0cm}{1.3cm}
$q'W^\pm\phi$  &  $41(17)$  &  $357(153)$  &  $2.17(1.37)$ &  \\ \hline
\rule[-0.6cm]{0cm}{1.3cm}
$q Z^0\phi$    &  $1(0)$  &  $0(0)$  &  $-(0)$ &  $120$ \\ \hline
\rule[-0.6cm]{0cm}{1.3cm}
$q \bar q\phi$ &  $1$  &  $0$  &  $-$ &  \\ \hline\hline
\rule[-0.6cm]{0cm}{1.3cm}
$q'W^\pm\phi$  &  $18(8)$  &  $300(128)$  &  $1.10(0.71)$ &  \\ \hline
\rule[-0.6cm]{0cm}{1.3cm}
$q Z^0\phi$    &  $0(0)$  &  $0(0)$  &  $0(0)$ &  $140$ \\ \hline
\rule[-0.6cm]{0cm}{1.3cm}
$q \bar q\phi$ &  $0$  &  $0$  &  $0$ &   \\ \hline
\multicolumn{5}{c}
{\rule{0cm}{.9cm}
{\Large Table IV}}  \\
\multicolumn{5}{c}
{\rule{0cm}{.9cm}}

\end{tabular}
\end{center}
\end{table}
\newpage
\thispagestyle{empty}

\begin{table}%[p]%[htbp]
\begin{center}
\begin{tabular}{|c|c|c|c|}
\hline
\rule[-0.6cm]{0cm}{1.3cm}
$S_{\rm tot}$  &  $B_{\rm tot}$  &  $S_{\rm tot}/\sqrt B_{\rm tot}$
&  $M_\phi$ (GeV) \\ \hline\hline
\rule[-0.6cm]{0cm}{1.3cm}
$120(44)$ &  $418(179)$  &  $5.87(3.32)$ &  $60$ \\ \hline
\rule[-0.6cm]{0cm}{1.3cm}
$84(33)$  &  $452(194)$  &  $3.95(2.37)$ &  $80$ \\ \hline
\rule[-0.6cm]{0cm}{1.3cm}
$64(26)$  &  $608(373)$ &  $2.60(1.35)$ &  $100$ \\ \hline
\rule[-0.6cm]{0cm}{1.3cm}
$43(18)$  &  $357(153)$  &  $2.28(1.46)$ &  $120$ \\ \hline
\rule[-0.6cm]{0cm}{1.3cm}
$19(8)$   &  $300(128)$  &  $1.10(0.71)$ &  $140$ \\ \hline
\multicolumn{4}{c}
{\rule{0cm}{.9cm}
{\Large Table V}}  \\
\multicolumn{4}{c}
{\rule{0cm}{.9cm}}

\end{tabular}
\end{center}
\end{table}

\vfill
\newpage
\thispagestyle{empty}

\begin{table}%[p]%[htbp]
\begin{center}
\begin{tabular}{|c|c|}     \hline
\rule[-0.6cm]{0cm}{1.3cm}
$\;\;\;\;\;$                     &
$\phi$                            \\ \hline
\rule[-0.6cm]{0cm}{1.3cm}
$W^\pm W^\mp$                    &
${M_{W^\pm}\over{s_W}}$     \\ %\hline
\rule[-0.6cm]{0cm}{1.3cm}
$Z^0Z^0$                         &
${M_{W^\pm}\over{s_Wc^2_W}}$     \\ \hline
\end{tabular}
\end{center}
\centerline{\Large Table VI}
\end{table}
\begin{table}%[p]%[htbp]
\begin{center}
\begin{tabular}{|c|c|c|c|c|}     \hline
\rule[-0.6cm]{0cm}{1.3cm}
$g$          &
$\gamma$     &
$Z^0$        &
$W^\pm$      &
$\phi$       \\ \hline
\rule[-0.6cm]{0cm}{1.3cm}
$(1,1)$                                                  &
$Q^q(1,1)$                                               &
${1\over{s_Wc_W}}(g_R^q,g_L^q)$                          &
${1\over{\sqrt2 s_W}}(0,1)$                              &
$\frac{m_q}{2M_{W^\pm}s_W}(1,1)$                         \\ \hline
\end{tabular}
\end{center}
\centerline{\Large Table VII}
\end{table}
\vspace*{\fill}

\end{document}